\documentclass[10pt,preprint1]{aastex}
\usepackage{graphicx}

\shorttitle{Tau Air-Shower Rates} \shortauthors{Fargion et al.}

\begin{document}

\title{ Tau Air Showers from  Earth}
\author{D. Fargion\altaffilmark{1,2}, P.G. De Sanctis Lucentini\altaffilmark{1}, M.
De Santis \altaffilmark{1}, M.Grossi \altaffilmark{1}}

\begin{abstract}
We estimate the  rate of observable Horizontal and Upward Tau
Air-Showers (HORTAUs, UPTAUS)
considering both the Earth opacity and the contribution of the
terrestrial atmosphere.
Our result applies to most neutrino telescope projects especially
to the EUSO space observatory. Using a compact analytical formula
we calculate the effective target volumes and masses for  Tau
air-showers emerging from the Earth.
The resulting model-independent effective masses for EUSO may
encompass - at $E_{\nu_{\tau}}\simeq  10^{19}$ eV - an average huge
volume ($\simeq 1020$ $km^3$) compared to current
neutrino experiments. Adopting simple  power law neutrino fluxes, $\frac{dN_{\nu}}{%
dE_{\nu}}$ $\propto$ $E^{-2}$ and $E^{-1}$, calibrated to GZK-like
and Z-Burst-like models,  we estimate that at $E \simeq 10^{19}$
eV nearly half a dozen horizontal shower events should be
detected by EUSO in three years of data collection considering
the 10\% duty cycle efficiency and a minimal $\nu_{\tau}$ flux
$\phi_{\nu} E_{\nu} \simeq 50 $eV cm$^{-2}$ s$^{-1}$ sr$^{-1}$.
The HORTAUS detection may test the "guaranteed" GZK neutrino flux
(secondaries of photopion production due to UHECR scattering onto
2.75 K cosmic background radiation). We also find that the
equivalent mass for an outer layer made of rock is dominant
compared to the water, contrary to simplified all-rock/all-water
Earth models and previous studies. Therefore we expect an
enhancement of neutrino detection along  continental shelves
nearby the highest mountain chains, also because of the better
geometrical acceptance for Earth skimming neutrinos. In this
picture, the Auger experiment might reveal such an increase at
$E_{\nu} \simeq 10^{18}$ eV (with 26 events in 3 yr) if the
angular resolution (both in azimuth and zenith) would reach an
accuracy of nearly one degree necessary to disentangle tau air
showers from common horizontal UHECR. Finally, we show that the
number of events increases at lower energies, therefore we
suggest  an extension of the EUSO sensitivity down to
$E_{\nu}\sim 10^{19}$ eV  or even below.

\end{abstract}

\affil{\altaffilmark{1} Physics Department,
Universit\'a  "La Sapienza", Pl.A.Moro ,\altaffilmark{2} INFN,
Rome, Italy}

\altaffiltext{1}{Physics Department, Universit\'a  "La Sapienza",
P.le A.Moro 5, 00185 Roma, Italy} \altaffiltext{2}{INFN Roma1,
Italy}

\

\section{Introduction: $\protect\tau$ air-showering from the Earth}

The study of ultrahigh energy upward and horizontal $\tau$ air
showers produced by $\tau$ neutrino interactions within the Earth
crust has been considered in recent years as an alternative way to
detect high energy neutrinos. The problem of $\tau$ neutrinos
crossing the Earth is indeed quite complicated because of the
complex terrestrial neutrino opacity at different energies and
angles of arrival. In addition, several factors have to be taken
into account, such as the amount of energy transferred in the
$\nu_{\tau}$ - $\tau$ lepton conversion, as well as the $\tau$
energy losses and interaction lengths at different energies and
materials. This makes the estimate of the links between the input
neutrino - output $\tau$ air shower very difficult. Such a
prediction is further complicated by the existence of a long list
of theoretical models for the incoming neutrino fluxes (GZK
neutrinos, Z-burst model flux, $E^{-2}$ flat spectra, AGN
neutrinos, topological defects). Many authors have investigated
this mechanism, however the results are varied, often in
contradiction among themselves, and the expected rates
range over a few order of magnitude (Fargion, Aiello, \&
Conversano 1999; Fargion 2002a; Feng et al. 2002; Bottai \&
Giurgola, hereafter BG03; Bugaev, Montaruli, \& Sokalski 2003;
Fargion 2003; Tseng et al. 2003; Jones et al. 2004; Yoshida et
al. 2004). So far, the majority of the current studies on this
topic is based on Monte-Carlo simulations assuming a particular
model of the incoming neutrino flux.

To face such a complex problem, we think that the simplest
approach is the best. First one has to disentangle the incoming
neutrino flux from the consequent $\tau$ air-shower physics.
Therefore, to establish the $\tau$ production rate we introduce
an effective volume and mass for Earth-skimming $\tau$'s, which
is independent on any incoming neutrino flux model. This volume
describes a strip within the Earth where
neutrino/antineutrino-nucleon, $\nu_{\tau} (\bar{\nu}_{\tau })
-N$, interactions may produce emerging $\tau^{-},\tau^{+}$
leptons which then shower in air.

We present a very simple analytical and numerical derivation (as
well as its more sophisticated extensions) which takes into
account, for any incoming angle, the main processes related to the
neutrinos and $\tau$ leptons propagation and the $\tau$ energy
losses within the Earth crust. Our numerical results are
constrained by upper and lower bounds derived in simple
approximations (enlisted in a final appendix). The effective
volumes and masses will be more severely reduced at high energy
because we are interested in the successful development of the
$\tau$ air-shower. Therefore we included as a further constraint
the role of the air dilution at high altitude, where $\tau$ decay
and the consequent air-shower may (or may not) take place.

We show that our results give an estimate of the $\tau$ air-shower
event rates that greatly exceeds earliest studies but they are
comparable or even below more recent predictions. To make our
derivation as simple as possible we present only the main formula
and plots, while full details of the calculations and of the
approximation limits will be discussed in the appendix.

We compare our general $\tau$ upward-going showers to  detectors
such as
the ongoing photo-fluorescence ground-based observatory Auger, we
present some definitive predictions for future projects like the
EUSO space observatory and we suggest how to enhance the neutrino
tau emergence especially at energies lower than $10^{19}$ eV.

The paper is divided in the following way: section \S 2 gives a
brief review of the expected Ultra High Energy Neutrino Sources.
Section \S 3 discusses the general air-shower neutrino telescope
scenario focusing on $\tau$ Neutrino Astronomy, with a minor hint
to the present and future underground $km^3$ neutrino detectors;
in section \S 4 we introduce the comparable horizontal Ultra High
Energy Cosmic Rays Air-Skimming events at high altitude (HIAS) and
the lower air-skimming neutrino induced air-shower; we calculate
the Earth opacity for neutrinos and leptons; we briefly remind the
lepton energy losses in different materials (water and rock), and
we present the effective Earth skin Volumes and Masses leading to
UPTAUs and HORTAUs. Such volumes and masses will be derived under
different approximations and experimental framework. In section \S
5 we present the event rate per km$^2$ assuming that the outer
layer of the Earth's crust is made either of  rock or of water and
we compare it with large area projects such as Auger and EUSO.
Section \S 6 will show our conclusions and suggestions for $\tau$
air-shower Neutrino Astronomy, mainly concentrating on the EUSO
experiment.


\section{Sources of High Energy Neutrinos}

\subsection{The atmospheric neutrinos background}

Cosmic Rays (CR) reach the Earth's atmosphere with a nearly
homogeneous and isotropic distribution with no astronomical memory
of their original place of birth. Parasite secondary neutrinos
originated by such blurred CR are abundantly hitting the
atmosphere, leading to an atmospheric neutrino background, with the
same nearly isotropic distribution as the parental CR flux in
celestial coordinates (with a power law spectrum $\propto$
$E^{-2.2}$). Their presence is a polluting signal for any future
Neutrino Astronomy. Nevertheless the atmospheric neutrino
background
is suppressed at high energy ($E_{\nu_{Atm}}> TeV$) and shows a spectrum ($%
\propto E^{-3.2}$) softer than that of the parental CR, because the
relativistic charged pion lifetime is larger than the time needed
to propagate through the atmosphere, and they do not easily decay
neither into muons nor into atmospheric neutrinos. Therefore the
detection of High Energy Neutrinos from galactic or extragalactic
objects is expected above $\sim 10$ TeVs (i.e. PeV - EeV - ZeV)
assuming a harder spectrum for neutrinos from astronomical sources.
It is worth noticing that the TeV-PeV charged cosmic rays are
confined within the Galaxy by the magnetic field, thus they are
more long-lived and tangled than direct galactic neutrinos; this
makes more likely (by at least two-three orders of magnitude) the
detection of atmospheric $\nu$'s (produced by galactic CR) compared
to those neutrinos from astrophysical sources, whose harder
spectrum may compensate this large difference.

At high energy ($> 10^{19}$ eV) CR losses are also of nuclear
nature (photo-pion and multi-pion productions) and Ultra High
Energy Cosmic Rays, UHECR, mainly nuclei and nucleons, often
produce charged pions leading to neutrino and gamma secondaries
nearby the original source, such as an Active Galactic Nucleus
(AGN). In this case we may expect an UHE neutrino astronomy (at
EeV-PeV) associated with such UHECR and with hard gamma photons ,
secondaries of neutral pions (Semikoz \& Sigl 2003). Such UHE
neutrinos are often labeled as AGN secondary neutrinos. While they
travel undeflected and unperturbed, the corresponding gamma are
often absorbed and degraded to GeV - MeV energies by intergalactic
backgrounds.

In the following sections we summarize the main sources of primary
UHE neutrinos such as the AGN, Gamma Ray Bursts (GRB) or more
exotic topological defects (TD), as well as the production of
secondary $\nu$'s by UHECR at GZK energies (GZK or cosmogenic
neutrino) and UHE neutrons. We also discuss the role of UHE
neutrinos as sources of UHECRs via $\nu - \nu_{relic}$ scattering,
as predicted by the Z-burst scenario.

\subsection{UHE $\protect\nu$ correlated to UHE neutrons}

Neutral EeV neutrons  may trace a corresponding neutrino imprint
through their beta decay in flight. Recently, a mild correlation
has been found between the Cosmic Ray excess in the AGASA data at
EeV energies with the EGRET gamma map towards Cygnus and the
Galactic Omega $17$ region (Hayashida et al. 1999; Fargion 2002a;
Fargion, Khlopov et al. 2003). The nature of such EeV galactic
anisotropy cannot be related to charged cosmic rays at high EeV
rigidity but it may be due
to EeV neutrons - able to avoid the galactic magnetic field bending and to survive up to galactic distances $%
D_{n}= 9.168\cdot \left(\frac{E_n}{10^{18} eV}\right)\cdot kpc $.
 These UHECR neutrons  anisotropy at $4\%$ level would imply an associated UHE neutrino flux.
Indeed neutrinos might be either the inescapable low tail of
neutron decay in flight, or the signature of UHE pions generated in
the same source with the UHE neutrons.
 In the first case (relic of neutron's beta decay in flight) the
 expected secondary  neutrino energy flux is a negligible ($\leq 1\%$) fraction of UHE
 neutron: $\phi_{\nu} \ll 12 $ eV $cm^{-2}s^{-2}sr^{-2}$ ; while in the
 latter case (assuming the equipartition of the energy into pion-neutron "in situ" production) the
 expected neutrino flux near EeV  might  exceed (up to  a factor
 ten) the UHE neutron flux $\phi_{\nu} \simeq 100 $ eV $cm^{-2}s^{-2}sr^{-2}$.

\subsection{UHE neutrinos from AGN and GRB}

UHECR escaping from a Jet of a Quasar (QSO) or an AGN, or a beamed
GRB - Supernova (SN) Jet may interact with the intense photon
fields generated by the source itself. These interactions lead to
the production of charged and neutral mesons fuelling a collinear
gamma and neutrino flux (Kalashev et al. 2002). The gamma rays and
the high energy cosmic rays may interact inside the much denser
environment of the source (AGN,GRB Jets) where the gamma opacity
suppresses most of the electromagnetic escaping signal. In this
scenario the main signal coming from the core of AGN might be
dominated by 'transparent' Ultra High Energy Neutrinos whose
fluxes may be even greater than the gammas. These AGN or GRB-SN
Jet neutrinos have been often modeled and predicted in the
PeV-EeV energy range, and they may be well correlated with gamma
photons produced by blazars, if they are originated by the
interactions of relativistic nuclei. EeV photons by neutral pion
decay are suppressed by photon-photon interactions and their
energy may be degraded to MeV-GeV energies. The neutrino energy
fluence may reach a value comparable or just below the one of the
diffuse gammas observed by EGRET: $\phi_{\nu} \simeq  10^3 $ eV
$cm^{-2}s^{-2}sr^{-2}$. The slope of the spectrum near the maximum
at PeV energy might be flat: $\phi_{\nu}$ $\propto E^{-2}$.




\subsection{UHE GZK neutrinos and the Z burst: neutrino masses imprint}

The origin of UHECRs with energies above a few times $10^{19}$ eV
is a phenomenon that has not yet been clearly understood. There
are nearly a hundred of such UHECR events, which surprisingly are
not clustered to any nearby AGN, QSRs or known GRBs within the
narrow volume (10-30 Mpc radius) defined by the cosmic 2.75
$K^{o}$ proton drag viscosity, the so-called GZK cut-off (Greisen
1966; Zatsepin \& Kuzmin 1966). Indeed the GZK cut-off implies the
shrinking of the UHECR propagation length to a few tens of Mpc,
making the expected UHECR astronomy a very local one. However
these events are neither correlated with any galactic or nearby
Local Group sources, nor with the Super-Galactic plane, nor with
nearby clusters. Following the Gamma Ray Burst lesson, the
observed UHECR overall isotropy is suggesting a very far away
cosmological origin. Such cosmic distances are not consistent
with the $< 10$ Mpc cut-off prescribed by the GZK effect.
Moreover, the presence of a few UHECR clustered events,
apparently correlated with a more distant, bright BL Lac
population makes this puzzle even more complicated. Such BL Lac
objects are in fact at far redshifts ($z> 0.1-0.3$) (Gorbunov,
Tinyakov, Tkachev, Troitsky 2002). This correlation and the
overall isotropy favor a cosmic origination for UHECRs, well
above the near GZK volume.

A light ($0.05$ eV $< m_{\nu} < 2$ eV ) relic neutrino may play a
role in solving the puzzle. Assuming a neutrino flux at
corresponding  energies ($2 \cdot 10^{21}$ eV $< E_{\nu} < 8 \cdot
10^{22}$ eV) ejected by distant BL Lacs, such UHE $\nu$'s may be
hitting relic light neutrinos clustered in Hot Dark Halos with a
characteristic size of a few Mpc (for instance as extended as the
Local Group) or of tens of Mpc (the supergalactic cluster
volume). This interaction produces an UHE Z boson (Z-Shower or Z
Burst model) whose secondary nucleons are the final observed
UHECRs on Earth. In synthesis, the UHE neutrino is the neutral
flying link (transparent to BBR photons) from BL Lac sources at
cosmic edges to our local universe, while the relic neutrinos,
possibly clustered in Mpc or ten Mpc volumes around the Local
Group, represent the target calorimeter.
In this scenario, a  neutrino mass (Dolgov 2002; Raffelt 2002)
fine-tuned at $m_{\nu} \simeq 0.4$ eV (Fargion, Grossi et al.
2000, 2001; Fodor, Katz, Ringwald 2002; Klapdor-Kleingrothaus et
al. 2001) or $m_{\nu} \simeq 0.1 \div 5$ eV,  may solve the GZK
paradox overcoming the proton opacity. (Fargion, Salis, 1997;
Fargion, Mele, \& Salis 1999; Yoshida et al 1998; Weiler 1999).
 In order to fit the UHECR  observed data a UHECR Z-burst neutrino
 flux  at $E \simeq  10^{19} eV $should  grow as $\phi_{\nu} \simeq 50 \cdot E^{-1}$ eV $cm^{-2}$  $s^{-1}$
 $sr^{-1}$.

Therefore, UHE neutrinos may induce, as primary particles, the
astronomy of the UHECRs and their detection may allow to identify
the sources where such UHE particles are produced. Yet,
independently on the exact GZK puzzle solution, the extragalactic
UHE nucleons produced in our GZK surrounding are
themselves source of UHE neutrinos $\nu_{\mu}$,$\bar{\nu_{\mu}}$$\nu_{e}$,$\bar{\nu_{e}}$%
, with energies $E \simeq 4\cdot 10^{19} eV $, and UHE gammas  by
photopion production.

However, the light neutrino mass required  to avoid the GZK paradox
does not solve the dark matter problem, although it is well
consistent with the  solar neutrino oscillation mass limits
(Anselmann et al. 1992; Fukuda 1998), with the most recent claims
of anti-neutrino disappearance by KamLAND (Eguchi et al. 2003) (in
agreement with a Large Mixing Angle neutrino model and $\Delta
{m_{\nu}}^2 \sim 7 \cdot 10^{-5}{eV}^2$), and with the atmospheric
neutrino mass splitting ($\Delta m_{\nu} \simeq 0.07$ eV). Finally,
a light neutrino mass in agreement with the Z-burst model (Fargion,
Grossi et al. 2000, 2001) may be compatible with the more recent
(but controversial) results from the neutrino double beta decay,
which sets the mass at $m_{\nu} \simeq 0.4$ eV
(Klapdor-Kleingrothaus 2002).







\subsection{Topological defects}

Current theories of particle physics predict that a variety of
topological defects have formed during the early stages of the
evolution of the universe. Such extremely heavy relic particles,
indicated as monopoles, strings, walls or necklaces, might be
originated as fossil remnants of phase transitions occurring at
the Grand Unified Theories energy scale of $10^{15}$ GeV. The
potential role of topological defects as an alternative
explanation to the origin of UHECRs above $10^{20}$ eV has been
proposed and investigated in the recent years (Bhattacharjee,
Hill, \& Schramm 1992; Sigl, Schramm, \& Bhattacharjee 1994).
 Their  decay lifetime is somehow fine-tuned with the Universe age.
 In addition,   pair annihilations of heavy particles (TD-like),  born in bounded binary
 system,  might also be a viable processes  in a narrow parameter
 range (Dubrovich, Fargion \& Khlopov 2004).
It is not clear if there are any sites in the universe capable to
accelerate the observed particles  at such high energies, while the
annihilation and decay of topological defects may produce prompt
extremely high energy cosmic rays and neutrinos. The search for
such neutrinos reaching the energy of GUT scale may shed light on
these models and theories. However, even if the TD scenario may
solve the puzzle of the particles' acceleration, the clustering of
some UHECR events (doublets or triplets) is hardly explicable by
the decay of a diffuse halo distribution of superheavy particles
(Uchihori et al. 2000, Tinyakov, \& Tkachev 2001). Such UHE
neutrino spectra from TD might follow a power law $\propto
E^{-3/2}$ in between the Z-burst $\propto E^{-1}$ and GZK $\propto
E^{-2}$ power laws.

\section{UHE Neutrino Telescopes:  Electron, Muon and Tau traces in underground ice-water and air detectors }

The very possible discover of an UHECR astronomy, the solution of
the GZK paradox, the very urgent rise of an UHE neutrino astronomy
are among the main goals of many new neutrino telescope projects.
Most detectors are related to the muon track in underground $Km ^3$
volumes, either in water or ice (Halzen, \& Hooper 2002). Such
detectors consist of a lattice of photo-multipliers spread over
large volumes and look for high energy neutrinos from the Cherenkov
emission of muons produced in $\nu - N $ interactions. The
detectors select preferentially upward-going muons entering from
below, originated by neutrinos propagating through the Earth, while
the vertical downward-going signals are polluted by atmospheric
muons. DUMAND and later BAIKAL detectors have been the pioneer
projects of under-ice detectors, more recently implemented by
AMANDA (Andres et al. 1997) and its extensions  AMANDA II at the
South Pole, with 677 photo-multipliers placed in the ice at depths
between $1500$ and $2000$ meters.


ANTARES (Montaruli et al. 2002) and NESTOR (Grieder et al. 2001) in
the Mediterranean sea, are the two ongoing projects for under-water
telescopes planned to be finished by 2006. Other detectors are
related to the UHE neutrino showering in air. The UHE neutrino
interactions  lead  to nuclear or electromagnetic showers. The UHE
neutrino-nuclei charged current interactions may produce electrons,
 muons or taus in air and/or water. In underground detectors the
electron-muon-tau interactions are leading to a Cherenkov flash
with or without a lepton track.
 Muon tracks are the most penetrating at  energies $\leq 10^{18} $ eV,
  while taus are more penetrating, as we shall see, at higher
  energies. The underground detectors are not  aimed at
  disentangling the lepton nature of the tracks.

 Among the experiments on the ground, Auger is designed to detect
air-showers both by muons bundles and their fluorescence lights
for downward  UHECR. Auger will investigate the region of the CR
spectrum around the GZK cut-off; however the characteristics of
the observatory are such that it may be able to detect also
HORTAUs events. The well amplified horizontal air-shower at a
slant depth larger than $2000$g $cm^{-2}$, (but not exceeding $\gg
10^4- 3\cdot 10^4$ g $cm^{-2}$), may trace an induced neutrino
air-shower event originated in air or, at higher rate, an
air-shower due to Tau decay in flight (HORTAUs) triggered by an
interaction within the nearby Andes mountain chain (Fargion,
Aiello, \& Conversano 1999; Fargion 2002a; Bertou et al. 2002) or
 emerging directly  from the Earth crust. We will discuss this
possibility in the following sections.

Auger is under construction in Argentina, it is planned to be
completed by 2005/2006 and it will be operative for about a decade.
It is the first prototype of a hybrid observatory for the highest
energy cosmic rays consisting of both a surface array of particle
detectors and fluorescence telescopes looking at the Cherenkhov
emission from high energy charged particles propagating through the
atmosphere. It will have an aperture of $7 \times 10^3$ km$^2$ sr,
which is roughly a factor 10 larger than HiRes. The observatory
includes also an analogue counterpart in the northern hemisphere
which will allow full sky coverage in order to study in detail the
spatial distribution of such events.

 Another competitive experiment is EUSO due to be
launched in a very near future. The EUSO detector is a wide angle
UV telescope that will be placed on the International Space
Station (ISS). It will look, in dark time, downward towards the
Earth atmosphere, and its aperture is such to cover a surface as
large as $\sim 1.6\times 10^{5}$ Km. Therefore it will encompass
AGASA-HIRES and Auger areas as well as their rate of UHECR events
(Cronin 2004). EUSO is designed to detect and measure
fluorescence and Cerenkov photons produced by the interactions of
UHECR in the atmosphere. As cosmic rays penetrate the atmosphere
they may originate both a photo-fluorescence signal (due to the
excitation of $N_{2}$ molecules) and a secondary (albedo)
reflection by Cherenkov photons. Given the large amount of UV
radiation emitted by the $N_{2}$ molecules, a substantial
fraction of UV photons is expected to reach the detector outside
the Earth's atmosphere at a height of $\sim 400$ km. For
instance, for a $10^{20}$ eV extended air shower (EAS), a few
thousand of photons are expected to reach EUSO . A much larger
number (at least two order of magnitude) of Cherenkov diffused
photons are expected for High Altitude Air-Showers (HIAS) and
HORTAUs. We will discuss in detail in the following sections the
possibility for EUSO to detect Horizontal Tau Air-Showers
originated within a very wide terrestrial skin volume around its
field of view (FOV)



As mentioned in section \S 2, the scattering of UHE neutrino onto
light relic neutrinos may solve the GZK paradox as predicted by the
Z-burst model. EUSO will play an important role to discriminate
between different scenarios proposed for the origin of UHECR and it
may discover UHE neutrino signals due to the Z burst or due to the
"guaranteed" GZK ones.

\section{ UHE $\protect\nu$ Astronomy by the Upward $\protect\tau$
Air-Showering}

While longest ${\mu}$ tracks in $km^3$ underground detector have
been, in last three decades, the main searched UHE neutrino signal,
tau air-showers by UHE neutrinos generated in mountain chains or
within the Earth skin crust at PeV up to GZK energies ($>10^{19}$
eV) have been recently proved to be a powerful amplifier in
Neutrino Astronomy (Fargion, Aiello, \& Conversano 1999; Fargion
2002a; Bertou et al. 2002; Hou Huang 2002; Feng et al 2002). Indeed
up-going UHE muons $E_{\mu} \geq 10^{14} $ eV, born from upward or
Earth skimming muon neutrinos, will propagate and escape from the
Earth atmosphere without any significant air-shower trace, $
L_{\mu}=6.6 \cdot10^{10} \,cm \left( \frac{E_{\mu}}{10^{14}\,eV}
\right)\gg  $ atmosphere height. The UHE electron  produced by UHE
$\nu_e$ will shower mainly within a very thin terrestrial crust and
its event will remain generally hidden and "buried" inside the
Earth. On the contrary UHE $\tau$ by UHE $\nu_{\tau}-N$
interactions  are able to cross thick earth crust layers and they
might emerge freely from the Earth. The   short  $\tau$ lifetime
will lead to a  decay in flight and to an amplified air-showering
at tuned $10^{14}-10^{19}$ eV energy band, within the main
interesting astrophysical range.

 Neutrino $\tau$ detectors searching for UPTAUs and HORTAUs will be (at least)
complementary to present and future, lower energy underground
$km^3$ telescope projects (such as AMANDA, Baikal, ANTARES, NESTOR,
NEMO, IceCube). In particular Horizontal Tau Air shower may be
naturally originated by UHE $\nu_{\tau}$ at GZK energies crossing
the Earth Crust just below the horizon  and/or nearby  high
mountain chains as the Andes. UHE $\nu_{\tau}$ are abundantly
produced by flavour oscillation and mixing from muonic (or
electronic) neutrinos, since galactic and cosmic distances are
larger than the neutrino oscillation lengths $
L_{\nu_{\mu}-\nu_{\tau}}=2.48 \cdot10^{19} \,cm \left(
 \frac{E_{\nu}}{10^{19}\,eV} \right) \left( \frac{\Delta m_{ij}^2
 }{(10^{-2} \,eV)^2} \right)^{-1} \simeq 8.3 pc.$   Therefore EUSO may observe HORTAUs events and it may set
constraint on models and fluxes, possibly answering some open
questions. Let us first describe the background UHECR events able
to mimic the HORTAUs signal that we are considering here.


\begin{figure}[htbp]
\par
\begin{center}
\includegraphics[width=12cm,height=8cm]{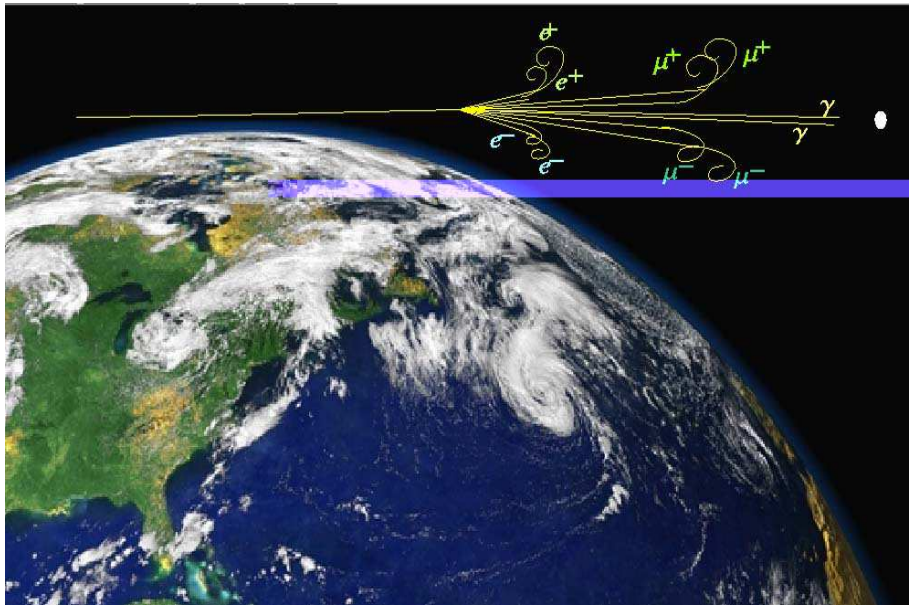}
\end{center}
\caption{A very schematic Horizontal High Altitude Shower (HIAS);
UHECRs (nucleon or nuclei or gamma) interact with the atoms of the
atmosphere at high altitude, producing a fan-shaped air-shower due
to the geo-magnetic bending of charged particles (mainly lepton
pairs) at high quota ($\sim 44 km$). The Shower may point to a
satellite as the old gamma GRO-BATSE detectors or to the more
recent Beppo-Sax, Integral, HETE, Chandra or the future ones as
Agile, Swift and GLAST (Fargion 2001b, 2001c, 2002a). These HIAS
Showers are extremely long (hundred km size) and they are often
split in five (or three) main components: $e^+ \, e^-,\protect\mu^+
\, \protect\mu^-, \protect\gamma $. Such multiple tails may be
detected in long horizontal air-showers by EUSO, and they may be
distinguished by their orthogonality to the local  magnetic fields.
At much lower altitude, below $10$ km height, there are very
similar UHE horizontal  showers due to neutrino-air interactions
leading to Neutrino Induced Air-Showers at  slant depth $X_{max}
\gg 10^3 g cm^{-2}$ } \label{fig1}
\end{figure}

\begin{figure}[htbp]
\includegraphics[width=8cm,height=4cm]{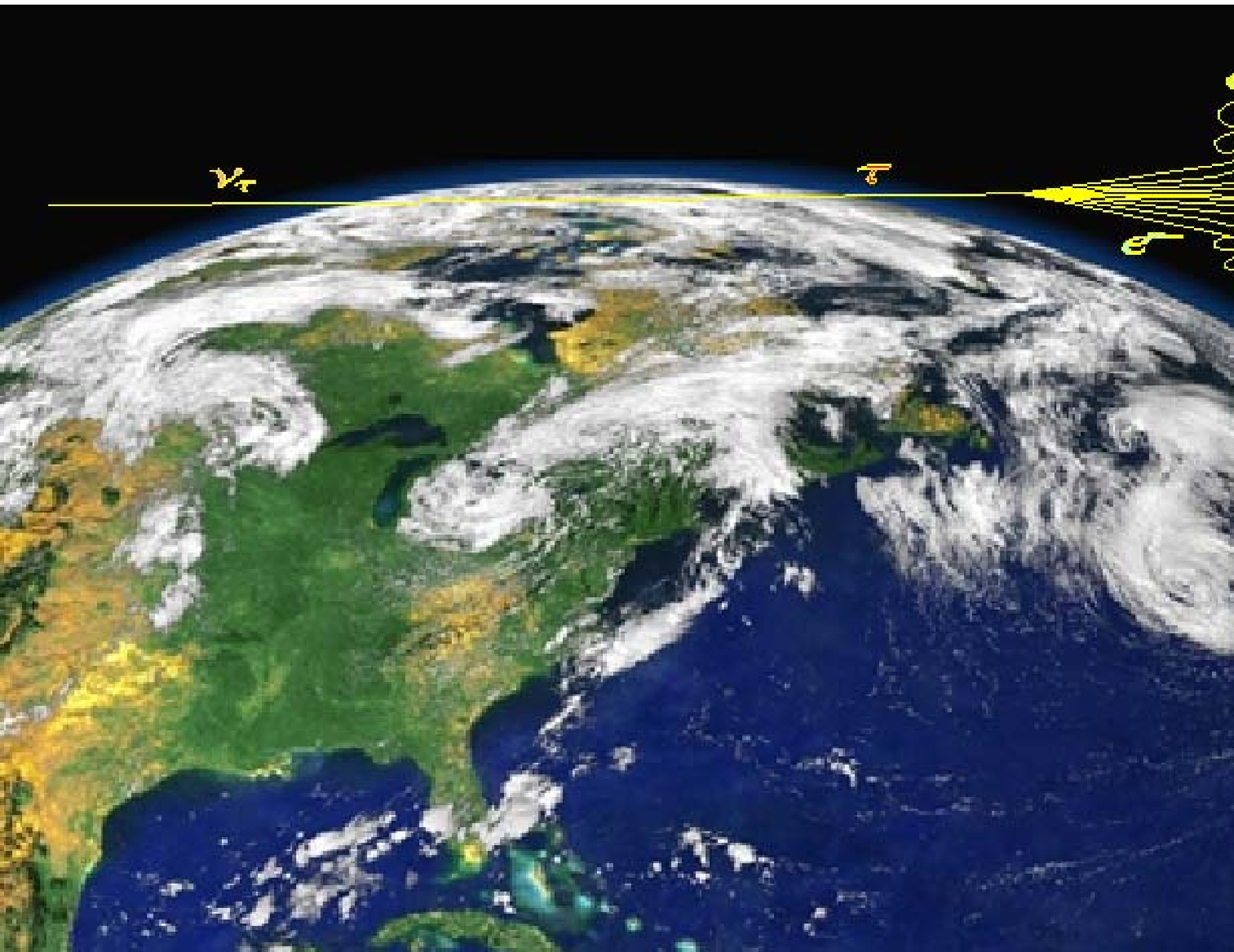} %
\includegraphics[width=8cm,height=4cm]{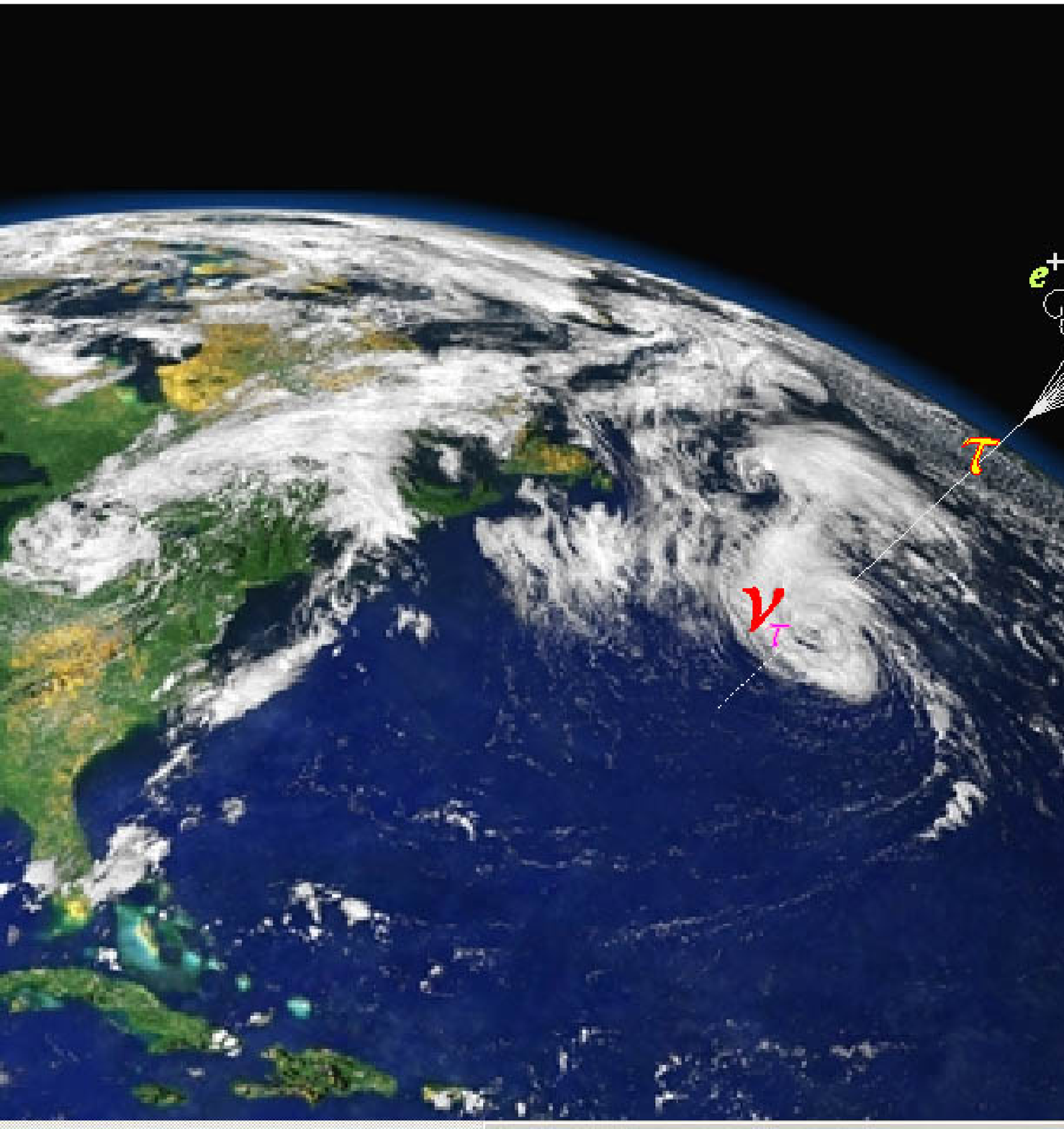} 
\caption{\textbf{left)} Horizontal Upward Tau Air-Shower (HORTAUS)
originated by UHE neutrino skimming the Earth: fan-shaped jets
arise because of the geo-magnetic bending of charged particles at
high quota ($\sim 23-40$ km). The shower signature may be
observable by EUSO just above the horizon. Because of the Earth
opacity most of the UPTAU events at angles $%
\protect\theta > 45-50^o$ will not be observable, since they will
not be contained within its current field of view (FOV).
\textbf{right)} A very schematic UPTAU air-shower at high altitude
($\sim 20-30 km$). The vertical  tail may be spread by geo-magnetic
field into a thin eight-shaped beam, observable by EUSO as a small
blazing oval (few dot-pixels) orthogonal to the local magnetic
field.} \label{fig3}
\end{figure}

  Figure \ref{fig1}   displays a characteristic scenario where a
  common UHECR hits horizontally the Earth atmosphere leading to a
  high altitude Air-Shower (HIAS, or Albedo  Shower).  This event takes
  place at an  altitude of $\sim 40$ km. A more interesting
  but similar  air-shower might originate at higher atmospheric density, i.e.
  at altitudes lower than $ 10$ km, by weak interacting neutrinos
  hitting the air nuclei. These events might be better detected by
  most present arrays or satellites at high energies $\sim
  10^{19}$eV.
  Figure \ref{fig3} describes the Horizontal Tau Air-Shower, HORTAUs
  event at  $\sim 10^{18} - 10^{19}$ eV  energy range below the horizon, whose detection efficiency
   is not contaminated by any downward  cosmic ray background.  Finally on the right-hand side
    of Figure \ref{fig3} we describe a much lower energy
 $\sim 10^{14} - 10^{17}$ eV  Upward  Tau  Air-Shower  UPTAUs
 induced as for HORTAUs by incoming tau neutrino able to cross the
 Earth inclined or almost vertical. At energies  $leq  2\cdot 10^{14}$ eV and below the UPTAUs
 more often take place  inside the terrestrial atmospheric layer.


\subsection{Effective Volume}

\begin{figure}[t]
\par
\begin{center}
\includegraphics[width=.50\textwidth]{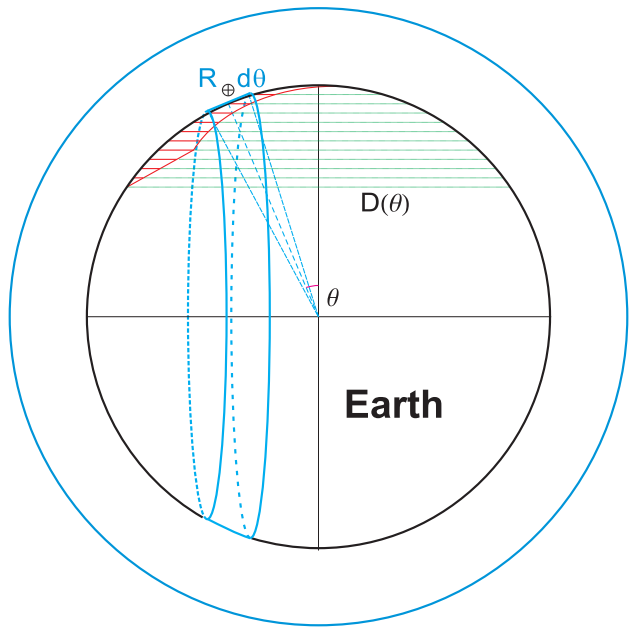}
\end{center}
\caption{A schematic representation of the possible incoming
neutrino trajectory through the Earth with the consequent
production of a tau lepton track. The thin shaded area on the left
(just below the Earth)  defines the volume where the interaction
occurs. This area might be thinner for $L_{\tau (\beta)}$ than for
$l_{\tau}$ interaction lenght. These emerging $\tau$'s may decay
in flight, producing the HORTAUs within the terrestrial
atmosphere.} \label{fig6}
\end{figure}

Let us define two main  effective volumes below the  Earth surface
where UPTAUs and HORTAUs might originate:

\begin{enumerate}
\item  A deeper skin volume   which is defined by all possible
$\nu_{\tau}$'s whose secondary $\tau$'s tracks escape  from the
Earth. This volume is obtained by considering the Earth opacity to
the incoming neutrinos and the inelasticity factor in the
conversion $\nu_{\tau} - \tau$, disregarding the exact outcoming
$\tau$ final energy. In this case the $\tau$ propagation (or
interaction) lenght, to be estimated in the following section,
$l_{\tau}$ is defined uniquely by the incoming neutrino energy
$E_{\nu_{\tau}}$ and by the energy losses in matter (see Fig.
\ref{fig6}).

\item  A thinner skin volume whose size may be defined by the
outcoming $\tau$ energy ( which is  related to the primary
$\nu_{\tau}$ energy). In this calculation we take into account the
earth's opacity, and the average energy losses of $\tau$ leptons
traveling through the earth's crust and escaping into the
atmosphere. The $\tau$ air-shower generation depends on the
presence of the  atmosphere, whose thickness defines an additional
suppression at highest energies. In this case the $\tau$
propagation (or interaction) lenght $L_{\tau (\beta)}$ (see next
section) is much shorter than the $l_{\tau}$ lenght (see Fig.
\ref{fig6}).

\end{enumerate}

Between these two scenarios, different effective volumes may be
obtained as a function of  the primary neutrino energy
$E_{\nu_{\tau_i}}$ or the newly born $\tau$ ($E_{{\tau}_i}$)  or
the final $\tau$ energy  ($E_{{\tau}_f}$) related to the observed
air-shower; such different possibilities  may lead to ambiguities
in the meaning and in the prediction of the  UPTAUs and HORTAUs
rates. Many authors calculate the UPTAUs rates as a function of
the incoming neutrino energy $E_{\nu_{\tau_i}}$. In the present
article we calculate the Volumes, Masses and Rates as a function
of  both the primary neutrino energy and  the final tau energy.
This procedure avoids  ambiguities and  allows an easier
comparison with other studies.

As we previously mentioned, our final effective volume should be
suppressed by a factor that includes the finite extension of the
earth's atmosphere (both for vertical and inclined air-showers)
which has never been considered before. This suppression guarantees
that one deals with only a fully developed air-shower and that
$\tau$ decays outside the terrestrial atmosphere are discarded.

\begin{figure}[htbp]
\begin{center}
\includegraphics[angle=0,width=12cm,height=8cm]{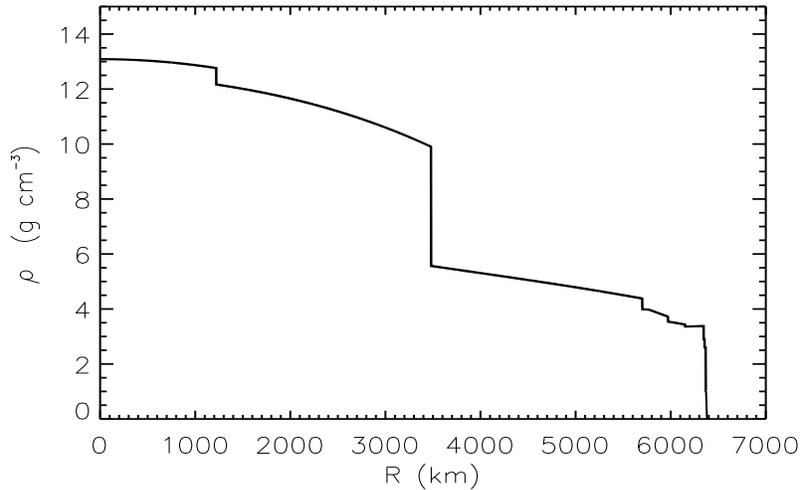}
\end{center}
\caption{The Inner terrestrial density profile inferred from the
Earth Preliminary Model.} \label{fig4}
\end{figure}

When one calculates the neutrino propagation through the Earth one
should take into account the complex internal structure of our
planet. Generally this may be approximated as a sphere consisting
of a number of layers with different densities, as represented in
the Preliminary Earth Model (Dziewonski 1989, see Fig. \ref{fig4}),
and for each density shell one should consider independent path
integrals. However, due to the complexity of these integrals we
shall skip them here, providing a more detailed calculation in the
Appendix.

\begin{figure}[htbp]
\begin{center}
\includegraphics[angle=270,scale=0.4]{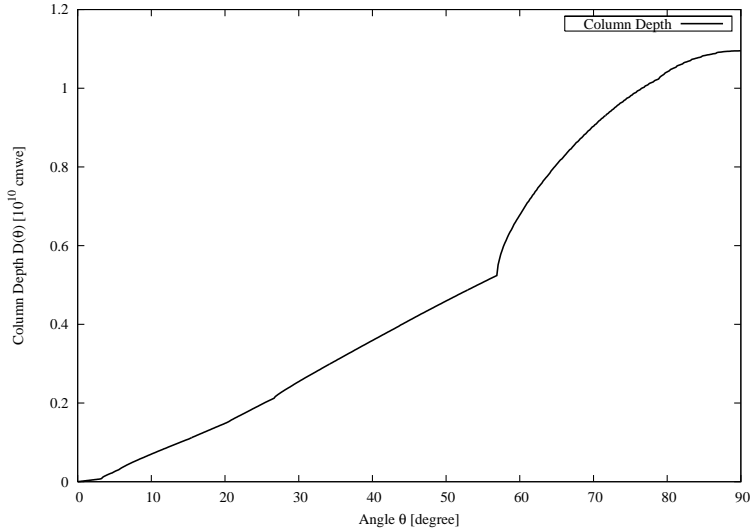}
\end{center}
\caption{Column depth as a function of the incoming angle having
assumed the multi-layers structure  given by the Earth
Preliminary Model.} \label{fig5}
\end{figure}

Alternatively for each chord described by a neutrino traversing the
Earth, one may introduce the column depth $D(\theta)$ defined as
$\int \rho(r) dl$, the integral of the density $\rho(r)$ of the
Earth along the neutrino path at a given angle $\theta$. The angle
$\theta$ is included between  the neutrino arrival direction and
the tangent plane to the earth at the observer location (see Fig.
\ref{fig6}) ($\theta = 0^o$ corresponds to a beam of neutrinos
tangential to the earth's surface) and it is complementary to the
nadir angle at the same location. The function $D(\theta)$ is
displayed in Fig. \ref{fig5}.

To calculate the effective volume we assume that the neutrino
traversing the Earth is transformed in a tau lepton at a depth $x$,
after having travelled for a distance $D(\theta) - x$. The
probability for the neutrino with energy $E_{\nu}$ to survive until
that distance is $e^{-(D(\theta) - x)/L_{\nu}}$, while the
probability for the tau to exit the Earth is $e^{- x/l_{\tau}}$. On
the other hand, as we will show in the next section, the
probability for the outcoming $\tau$ to emerge from the Earth
keeping its primary energy $E_{\tau_i}$ is $e^{- x/L_{\tau
(\beta)}}$. By the interaction length $\L_{\nu}$ we mean the
characteristic length for neutrino interaction; as we know its
value may be associated to the inverse of the total
cross-section  $\sigma_{Tot}= \sigma_{CC} + \sigma_{NC}$, including
both charged and neutral current interactions. It is possible to
show that using the $\sigma_{CC}$ in the $e^{-(D(\theta) -
x)/L_{\nu}}$ factor includes most of the ${\nu}_{\tau}$
regeneration along the neutrino trajectory making simpler the
mathematical approach. Indeed the use of the total cross-section in
the opacity factor above  must be corrected  by the
multi-scattering events (a neutral current interaction first
followed by a charged current one later); these additional $relay$
events ("regenerated taus") are summarized by the less suppressing
$\sigma_{CC}$ factor in the $e^{-(D(\theta) - x)/L_{\nu CC }}$
opacity term. The only difference between the real case and our
very accurate approximation is that we are neglecting a marginal
energy degradation (by a factor $0.8$) for only those " regenerated
taus " which experienced  a previous neutral current scattering. At
energy below $10^{17}$ eV there is also a minor
 $\nu_{\tau}$ regeneration (absent in $\mu$ or $e$
 case) that may be neglected because of its marginal role in the
 range of energy ($ >> 10^{17}$ eV)  we are interested in.

 The effective volume is given by

\begin{equation}
\frac{V_{Tot}(E_{\nu})}{A}=\int^{\frac{\pi}{2}}_{0} \int^{D(\theta)}_{0} e^{-%
\frac{D(\theta) -x}{L_{\nu_{{CC}}}(E_\nu)}}e^{\frac{-x}{%
l_{\tau}(E_{\tau})}}\sin{\theta}\cos{\theta}d\theta dx
\end{equation}

Under the assumption that the $x$ depth is independent of $L_{\nu}$ and $%
l_{\tau}$, the above integral becomes:

\begin{equation}
\frac{V_{Tot}(E_{\nu})}{A}=\left(\frac{l_\tau}{1-\frac{l_\tau}{L_\nu}}%
\right)\int^{\frac{\pi}{2}}_{0} \left(e^{-\frac{D(\theta)}{L_{\nu CC
}(E_{\nu}))}}-e^{-{\frac{D(\theta)}{l_\tau (E_{\tau})}}}\right) \sin{\theta}%
\cos{\theta}d\theta
\end{equation}

Given that $e^{-{\frac{D(\theta)}{l_\tau}}} \ll e^{-\frac{D(\theta)}{L_{\nu
CC}}}$, the second exponential in the integral may be neglected and the
relation can be rewritten as

\begin{equation}
\frac{V_{Tot}(E_{\tau})}{A}=\left(\frac{l_{\tau}(E_\tau)}{1-\frac{%
l_{\tau}(E_{\tau})}{L_{\nu}(\eta E_\tau)}}\right)\int^{\frac{\pi}{2}}_{0}e^{-%
\frac{D(\theta)}{L_{\nu_{CC}}(\eta
E_\tau)}}\sin{\theta}\cos{\theta}d\theta \label{eq_vol_ltau}
\end{equation}

where the energy of the neutrino $E_{\nu}$ has been expressed as a function
of $E_{\tau}$ via the introduction of the parameter $\eta = E_{\nu}/E_{\tau_f}$%
, the fraction of energy transferred from the neutrino to the
lepton. At energies greater than $10^{15}$ eV, when all mechanisms
of energy loss are neglected, $\eta = E_{\nu}/E_{\tau_f} =
E_{\nu}/E_{\tau_i} \simeq 1.2$, meaning that the 80 \% of the
energy of the incoming neutrino is transferred to the newly born
$\tau$ after the $\nu - N$ scattering (Gandhi 1996, 1998).

When the energy losses are taken into account, the final $\tau$
energy $E_{\tau_f}$ is a fraction of the one at its birth,
$E_{\tau_i}$. Their ratio $x_i = E_{\tau_f} / E_{\tau_i}$ is
related to $\eta$ by the following expression

\[
\eta(E_{\tau_f}) = \frac{E_{\nu}}{E_{\tau_f}} =
\frac{E_{\nu}}{E_{\tau_i}} \frac{E_{\tau_i}}{E_{\tau_f}} \simeq
\frac{1.2}{x_i(E_{\tau_f})} .
\]



Once the effective volume is found, we introduce an effective mass
defined as

\begin{equation}
\frac{M_{Tot}}{A}=\rho_{out}\frac{V_{Tot}}{A}
\label{eq_M_ltau}
\end{equation}

where $\rho_{out}$ is the density of the outer layer of the Earth crust: $\rho_{out}=1.02$ (water) and $%
2.65$ (rock). Before showing the effective Volume and Mass in
Figure \ref{fig7} for different densities we discuss the $\tau$
lepton energy losses needed to estimate the above formula.



\subsection{Neutrino and Tau Interactions and Energy losses: the beta
function}

Ultrahigh-energy tau neutrinos may be detected by observing tau
air showers originated in the $\nu -$$N$ interactions as a
neutrino beam crosses the Earth. The tau lepton energy is the only
observable quantity we can measure, therefore we have to determine
the relation between the energy of the primary neutrino and of the
outgoing lepton $\tau$, $E_{\nu_i} = E(E_{\tau_f})$. Moreover, one
has to consider that $\tau$ leptons traveling across the Earth
lose energy while interacting with the nucleons. At low energies
the relation between $E_{\nu}$ and $E_{\tau}$ is easy to define,
and it may be approximated as $E_{\tau_i} = (1 - y) E_{\nu}$ where
$y \sim 0.2$ is the fractional energy transferred to the nucleus.
The  elasticity parameters are different for the $\nu$ and
$\overline{\nu}$. While at low energy they differ by nearly a
factor $2$, at highest energy $y$ converges to the same $0.8$
limit for both neutrinos. We use their average value in our
analysis.

\begin{figure}[htbp]
\begin{center}
\includegraphics[angle=270,scale=0.4]{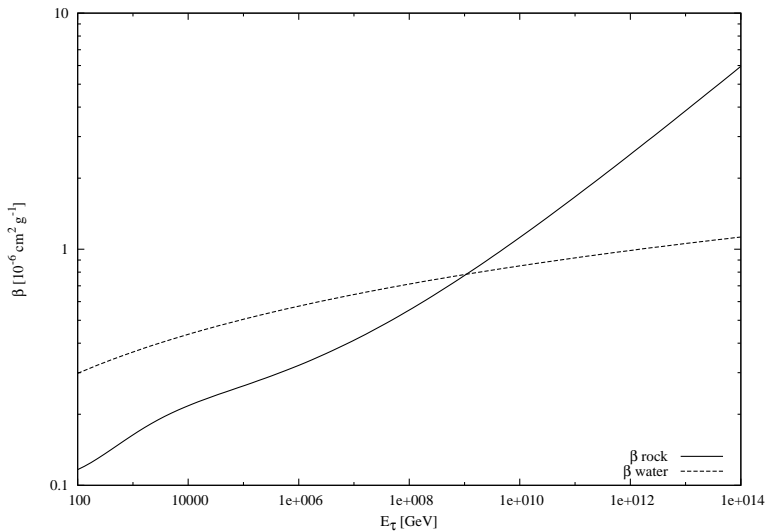}
\end{center}
\caption{Beta energy values in water ($\protect\rho =$1) and rock ($\protect%
\rho = 2.65$)} \label{beta}
\end{figure}

Increasing the  energy, different mechanisms  suppress the lepton's
propagation: first ionization, then bremsstrahlung, pair
production and (at $E>10^8 GeV$) photo-nuclear reactions dominate our calculation of the emerging final lepton energy $%
E_{\tau_f}$. The energy losses are described by the equation

\begin{equation}
- \frac{dE}{dX} = \alpha + \beta E
\end{equation}

where $\alpha$, the ionization energy loss, is negligible at
energies above several hundreds GeV and $\beta = \sum_i \beta_i$
is the sum of the remaining three mechanisms, each one denoted by
the index $i$. For tau leptons the photonuclear and pair
production processes are more important compared to
bremsstrahlung (Dutta et al. 2001), but above $10^{5}$ GeV the
photonuclear interactions become the most efficient mechanism of
energy loss. The $\beta$ value is weakly dependent on the energy
and for small slant depth $dX$ it may be assumed as constant.
Thus, the final energy of the tau becomes $E_{\tau_f} =
E_{\tau_i} e^{-\beta (E_{\tau_i}) dX}$, where ($ \rho
\beta^{-1}$) defines a scale-length where the energy of the
leptons is not severely suppressed.  Fig. \ref{beta} displays the
$\beta$ dependence on the tau energy for leptons propagating
through rock and water, including all the three components
(bremsstrahlung, pair production and photonuclear interactions)
whose expression has been derived by Dutta et al. (2001) for the
rock and by Jones et al. (2004) for the water. Here both
functions have been extrapolated to higher energies.


In Fig. \ref{l_tau} we show the tau interaction length for
different matter densities. When energy losses are negligible the
tau range is given by the tau decay length $R_{\tau}$. Including
the energy losses the tau propagation is suppressed compared to the
free decay length. The $\tau$ interaction length - $l_{\tau}$ - for
escaping leptons from the Earth has been derived with a convenient
approximation of the solution of the following pair of transcendent
equations (see Fargion 2002a for a more detailed description)

\begin{equation}
\frac{\ln (1/x_i)}{\rho_r [\beta_{0 \tau} + \beta_{1 \tau} \ln
(E_{\tau_i} / E_{0 \tau})]} = 492 x_i \frac{E_{\tau_i}}{E_{0 \tau}}
\label{l_tau_beta_i}
\end{equation}

\begin{equation}
\frac{\ln (1/x_i)}{\rho_r [\beta_{0 \tau} + \beta_{1 \tau} \ln
(E_{\tau_i} x_i / E_{0 \tau})]} = 492 x_i \frac{E_{\tau_i}}{E_{0
\tau}} \label{l_tau_beta_f}
\end{equation}

where $x_i = (E_{\tau_f} / E_{\tau_i})$,  $E_{0 \tau} = 10^{14}$
eV,  492 is the tau length in cm at this energy , and $\beta_{0
\tau}$ contains comparably constant terms from pair production and
photo-nuclear interactions, while $\beta_{1 \tau}$ is mainly due to
photo-nuclear effects.

\begin{figure}[h]
\begin{center}
\includegraphics[angle=0,scale=0.6]{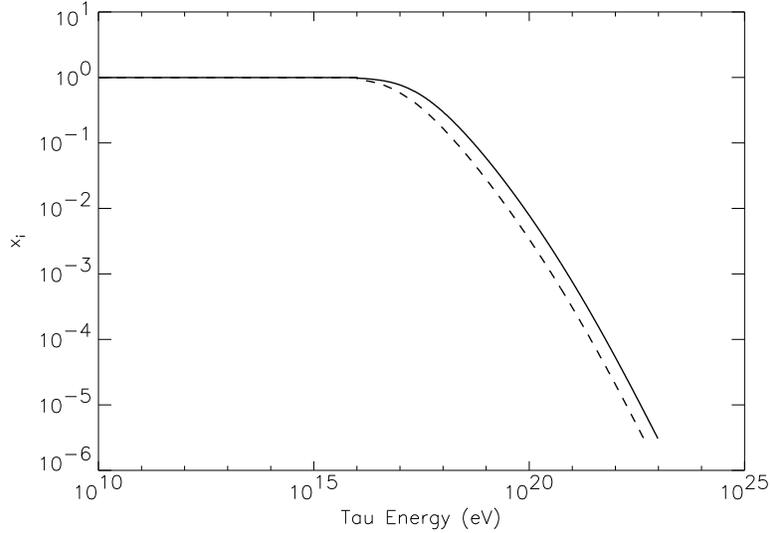} %
\caption{ The ratio between the final and initial $\tau$ energy
ratio in water (solid line) and in  rock.} \label{x_i}
\end{center}
\end{figure}

For instance we have used as a convenient expression for the water
$\beta_{0 \tau} = 2.3 \cdot 10^{-7}$ cm$^{2}$ g$^{-1}$ and
$\beta_{1 \tau} = 4.5 \cdot 10^{-8}$ cm$^{2}$ g$^{-1}$ (see Fargion
2002a). However in this paper we use the exact behaviour of $\beta$
as a function of $E_{\tau}$, as it is displayed in Fig. \ref{l_tau}
containing all the terms contributing to the energy losses
respectively for water and rock. The $\tau$ interaction length is
then given by

\begin{equation}
l_{\tau} = 492 x_i \frac{E_{\tau i}}{E_{0 \tau}} \: cm
\label{l_tau_2002}
\end{equation}
 where the  $x_i$, displayed in Fig. \ref{x_i} as a function
 of the outcoming $\tau$ energy, is the average value found  at any energy by the set of two
 independent transcendental  equations above.

At highest energies, above 10$^{10}$ GeV weak interactions play a
role in the energy losses of the $\tau$. The interaction length of
the charged-current $\tau N$ process is shorter than the tau decay
length above  10$^{10}$ GeV but already the energy losses
length-scale (mainly due to nuclear losses ) suppress the tau
propagation at earlier high energies. In Fig. \ref{l_tau}
(left-hand side) this interaction length is indicated as $L_{\nu
CC}$, since one can show that the cross section of the electro-weak
reaction $\tau - N$ is equivalent to the one of the $\nu - N$
scattering (Fargion 2002a). These three processes combine into the
curve $l_{\tau}$ shown in Fig. \ref{l_tau} that we have calculated
for both water and rock.

Finally, in the righthand side of Fig. \ref{l_tau} we compare the
value of $ l_{\tau}$ with the scale-length defined by $L_{\tau
(\beta)}$, that we will use for our calculation of the inner
effective volume for tau surviving with most primary energy. This
interaction length is defined as

\begin{equation}
L_{\tau (\beta)} = \left( \beta \rho + \frac{1}{L_{\nu_{CC}}} +
\frac{1}{R_{\tau}} \right)^{-1} \label{l_tau_beta}
\end{equation}

Again it is displayed for a $\tau$ propagating through water
(dash line) and rock (dotted line) and it is compared with
$l_{\tau}$. Compared to $l_{\tau}$, the interaction $L_{\tau
\beta}$ is shorter at energies greater than few times $10^{17}$
eV, therefore it defines a thinner effective volume. Within such
interaction volume the outcoming $\tau$ preserves most of the
initial energy determining a harder tau air-shower spectrum.
 This procedure gives results equivalent to those of Monte-Carlo predictions
  which we did verify but are not displayed here.

\begin{figure}[htbp]
\includegraphics[angle=270,scale=0.3]{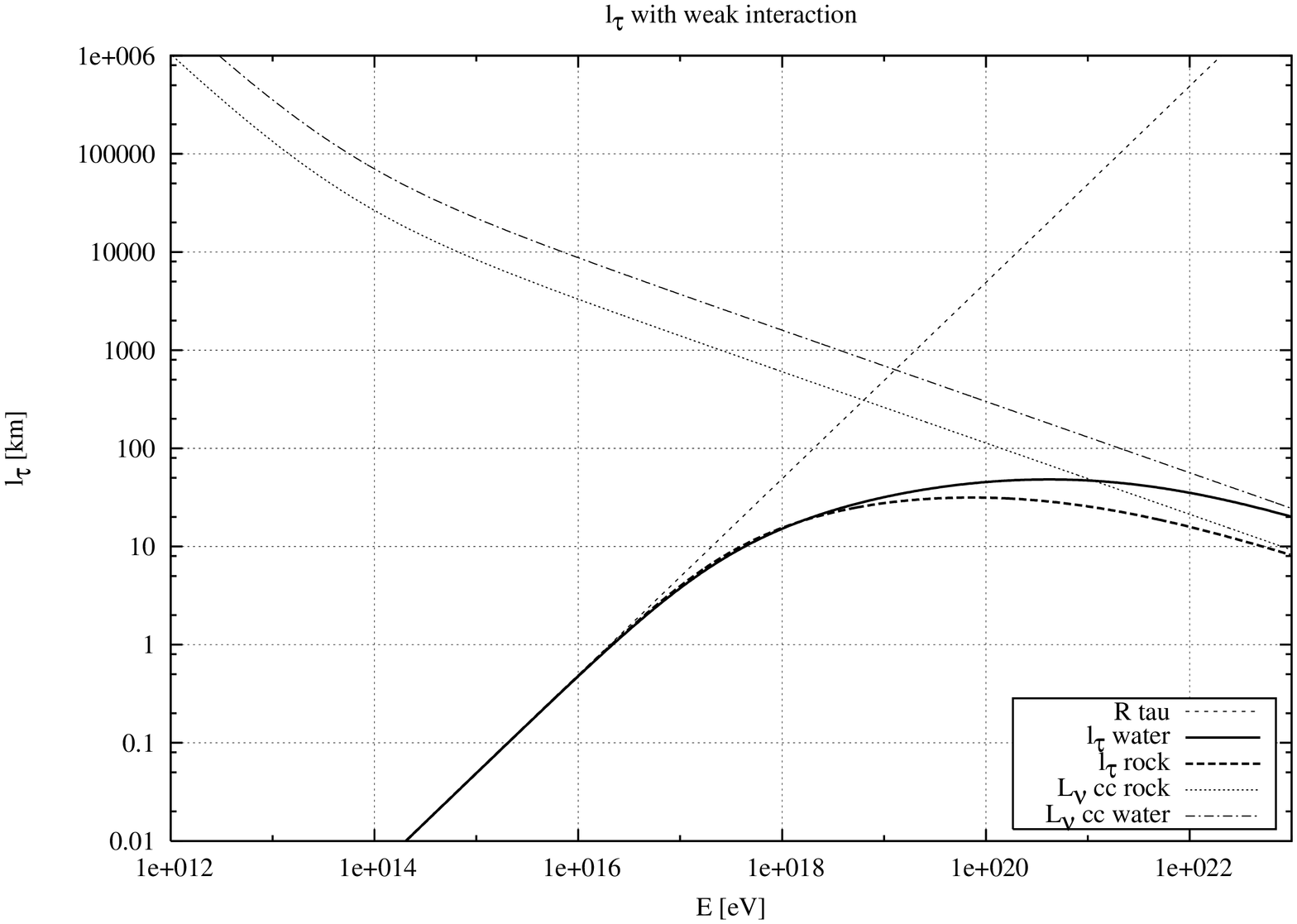} %
\includegraphics[angle=270,scale=0.3]{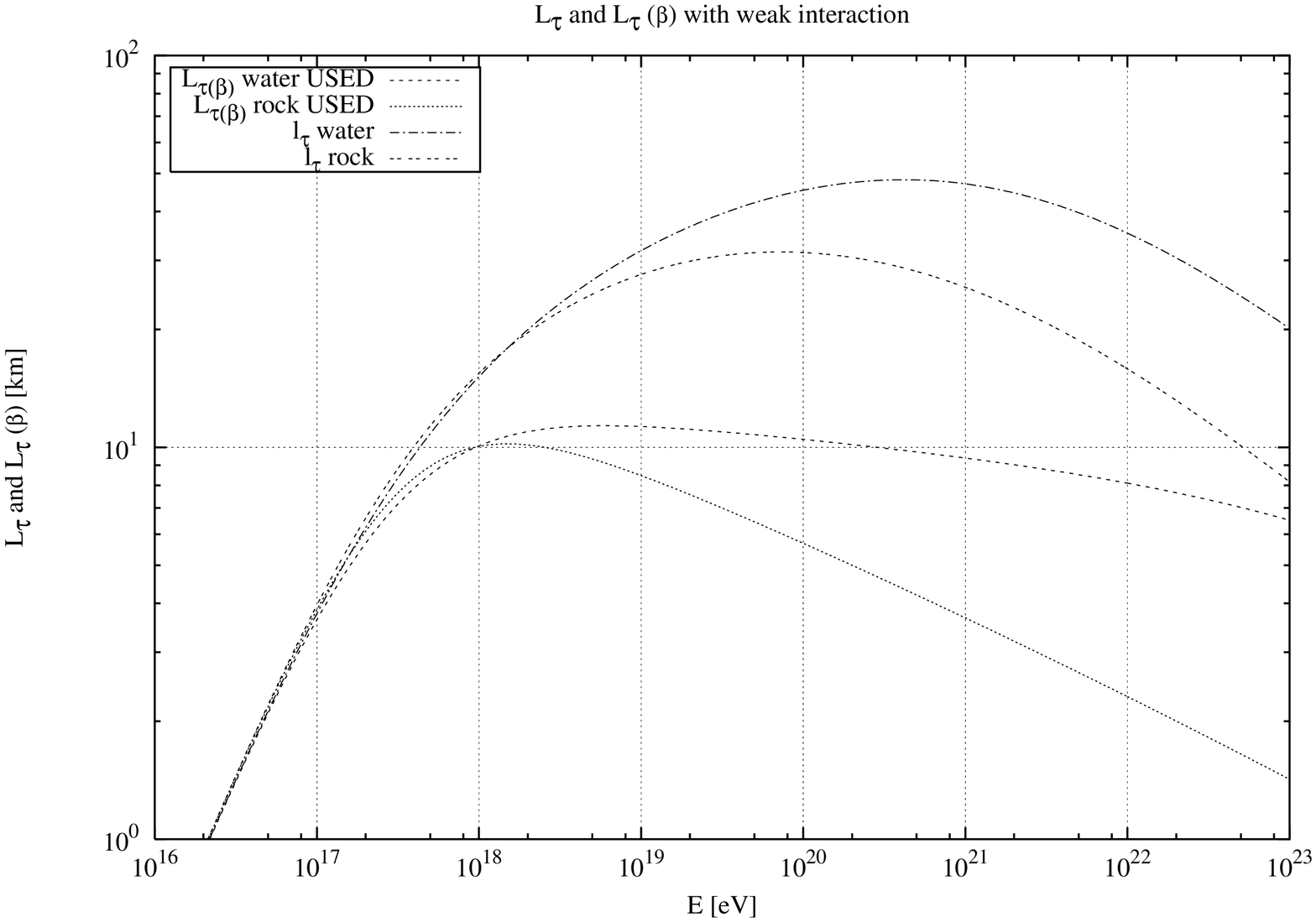}
\caption{\textbf{left)} Lepton $\protect\tau$ interaction Lengths
for
different matter densities: $R_{\protect\tau} = c\cdot {\protect\tau_{%
\protect\tau}}\cdot {\protect\gamma_{\protect\tau} } $ is the free $\protect%
\tau$ range (dashed line),
$l_{\tau}$ is the $\protect\tau$ propagation length in water (solid
line) and rock (thick dashed line), including all known
interactions and energy losses. $L_{\nu CC}$ is the $\tau$
propagation length due to electro-weak interactions with nucleons,
($\tau - N$), and it is equivalent to that of the $\nu - N$
scattering (Fargion 2002a). Again this length is displayed for both
water (dashed-dotted line) and rock (dotted line) densities.
 \textbf{right)} Comparison between $L_{\tau (\beta)}$ and $l_{\tau}$ for rock and water. The latter curves ($l_{\tau}$)
 are identical (but in different scale) to those shown on the left-hand side.  As one
can see from the picture, $L_{\tau (\beta)}$ is shorter than $l_{\tau}$ at energies above $10^{17}$ eV, thus it
 corresponds to a smaller effective volume where
$\protect\tau$'s are produced while keeping most of the primary
neutrino energy. The energy  label on the x axis refers to the
newly born tau.} \label{l_tau}
\end{figure}

\subsection{Final  $\tau$ skin effective volume in different detectors}

\begin{figure}[tp]
\begin{center}
\includegraphics[angle=270,scale=0.3]{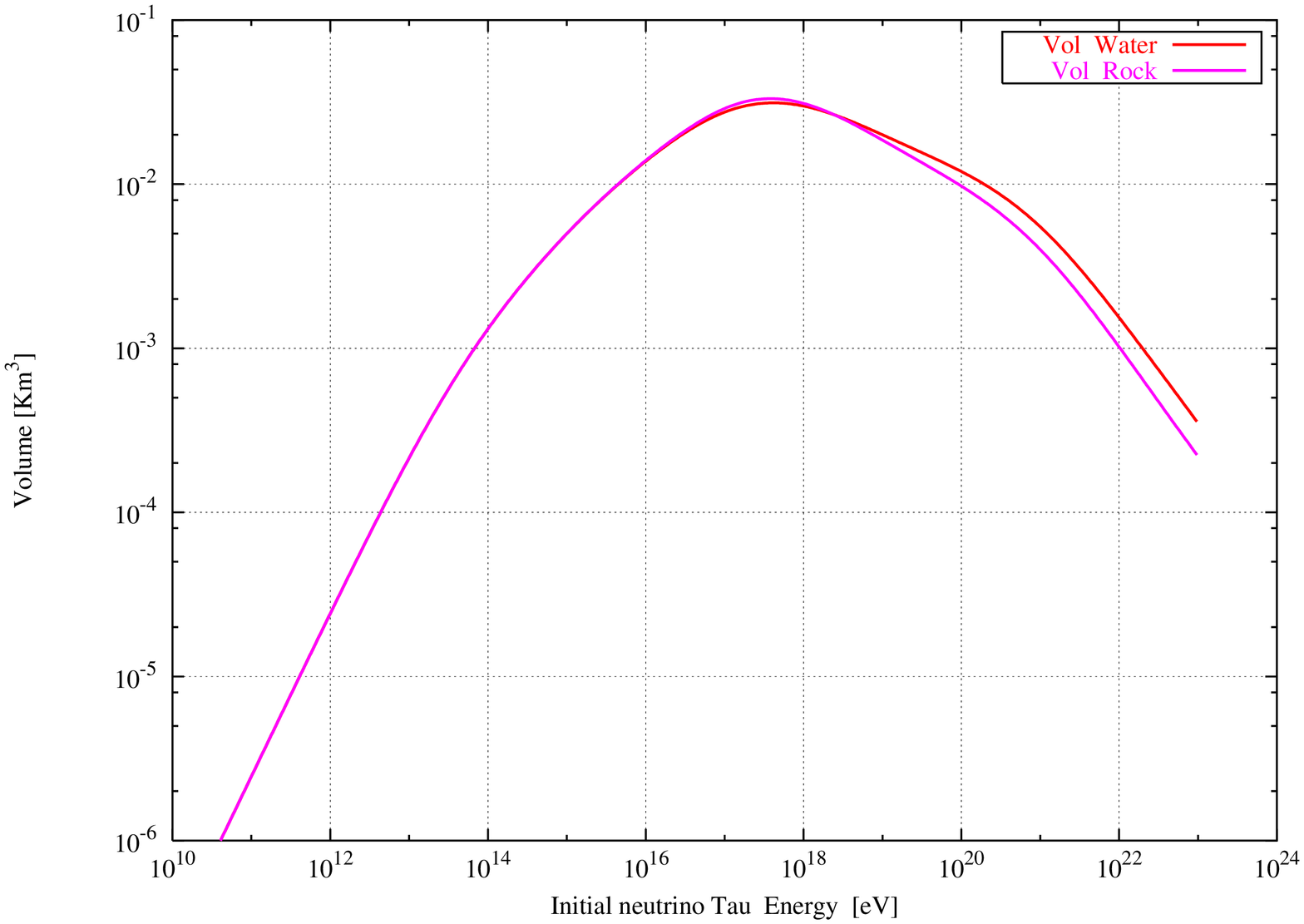}
\includegraphics[angle=270,scale=0.3]{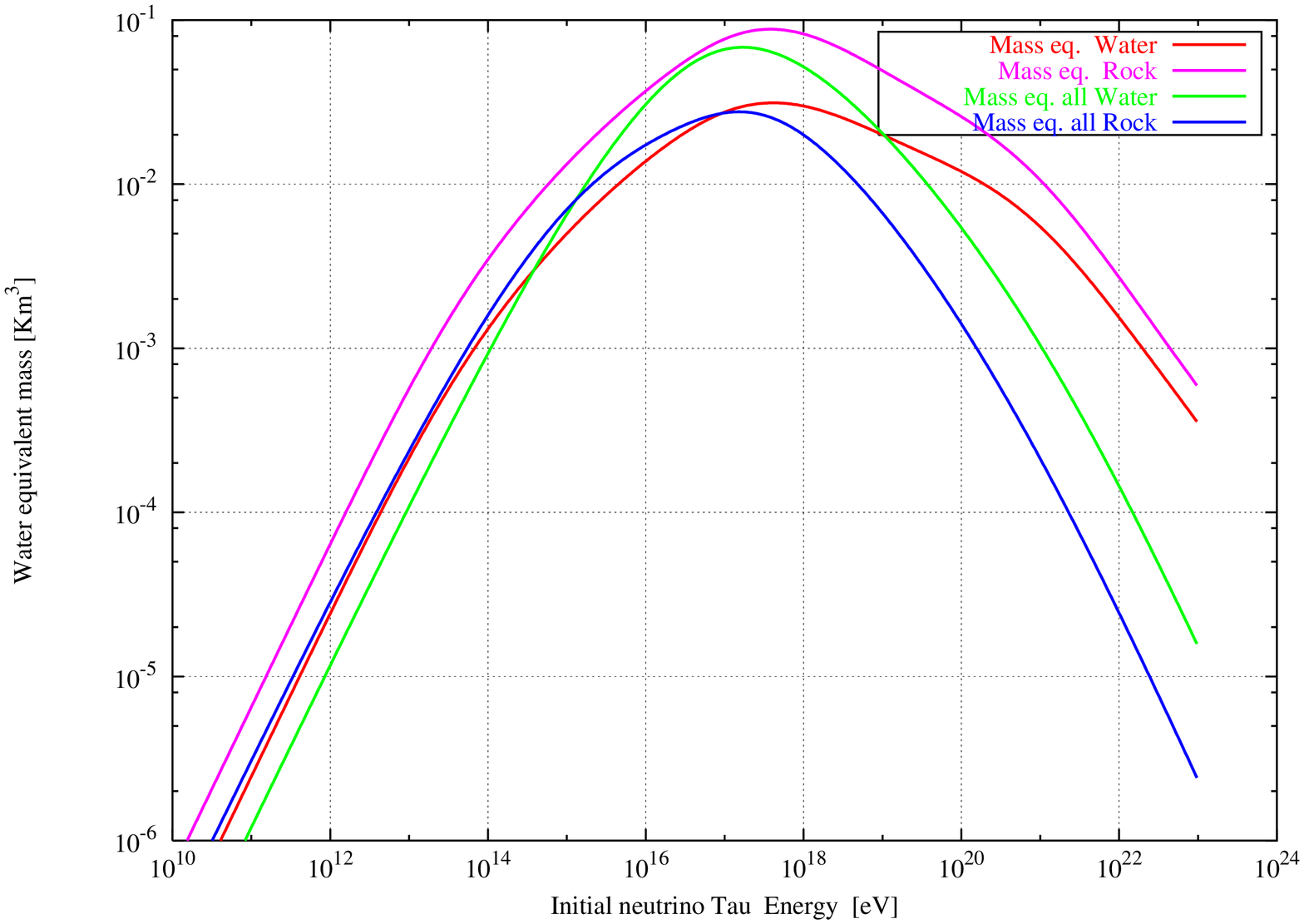}
\end{center}
\caption{\textbf{left)} The effective Volume for UPTAUs and HORTAUs
for a detector acceptance of 1 km$^2$ having neglected the role of
the atmospheric layer in the development of the tau air shower. We
have chosen $\eta = 1.2$, implying that the 80 \% of the initial
neutrino energy has been transferred to the newly born $\tau$. The
volume has been estimated as in Eq. \ref{eq_vol_ltau} using
$l_{\tau}$ as interaction length. \textbf{right)} The corresponding
effective Mass for UPTAUs and HORTAUs per km$^2$ (see Eq.
\ref{eq_M_ltau}) with the same approximations used for the
calculation of the volume. The two additional curves represent a
simplified model of the Earth treated as an unique homogeneous
sphere of water ($\rho =$1) or of rock ($\rho = 2.65$). The
horizontal line at $M = 10^{-2}$ $km^3$ corresponds to the air mass
above the same $km^2$  area. Note that the masses for an Earth made
entirely of water or rock has been obtained under the assumption of
a finite thickness for the terrestrial atmosphere (an horizontal
extension of $600$ km). The same constraint will be applied to the
final calculation of the effective volume and mass.} \label{fig7}
\end{figure}

Having derived the relation between $E_{\nu_i}$ and $E_{\tau_f}$,
and having defined a length scale for the tau energy losses, we can
calculate the $\tau$ skin volume as a function of the $\nu$ as well
as $\tau$ energy. We remind to the reader that $V_{eff}$ is the
effective volume where Ultra High Energy neutrinos interactions
within the Earth lead to UHE Taus. In this section we introduce the
complete calculation and we show the plots of the corresponding
effective volume and mass for the Earth preliminary model. In Fig
\ref{fig7} we are showing the effective volume $V_{eff}(E_{\nu_i})$
and its consequent mass $M_{eff}(E_{\nu_i})$ for a detection
acceptance of $A= 1 km^2$, considering the simplest case where we
include all the $\tau$ events regardless of the exact $\tau$ final
energy and neglecting the $\tau$ air-shower occurrence.

At $10^{19}$ eV we find an effective Volume $V_{eff} = 2\cdot
10^{-2}$ km$^3$  and a mass $M_{eff} = 2\cdot 10^{-2}$ km$^3$
(water equivalent) assuming an outer layer of the Earth made of
water ($\rho = 1.02$). For the rock instead we obtain $V_{eff} =
1.86 \cdot 10^{-2}$ km$^3$, and $M_{eff} = 4.95 \cdot 10^{-2}$
Km$^3$ (water equivalent).
  This simple result states the dominant role of UPTAUs and HORTAUs
   volume and masses   compared to the downward induced neutrino events  in air: $V_{air} = 10^{-2}$ km$^3$, $M_{eff} = 10^{-2}$ Km$^3$ (water equivalent)
  in the wide energy window $ 10^{16} eV< E_{\nu_i}< 10^{20} eV $
  almost un-affected by any atmospheric neutrino background.

We first present the calculation of effective mass and volume in
two cases: first for a generic detector with an acceptance of one
km square unit area, then we present the same calculation for the
characteristics of the EUSO experiment. Finally we compare EUSO
with the Auger observatory.

The expression of the effective volume in the most general case,  using the
Earth preliminary model is given by

\begin{equation}
\frac{V_{Tot}(E_{\tau})}{A}=\left(\frac{L_{\tau (\beta)}(E_\tau)}{1-\frac{%
L_{\tau (\beta)}(E_{\tau})}{L_{\nu_{CC}}(\eta E_\tau)}}\right)\int^{\frac{\pi%
}{2}}_{0}e^{-\frac{D(\theta)}{L_{\nu_{CC}}(\eta E_\tau)}}\sin{\theta}\cos{%
\theta}d\theta  \label{Veff_lbeta}
\end{equation}

where $L_{\tau \beta}$ is defined by Eq. \ref{l_tau_beta}.  This
interaction length (shorter than  $l_{\tau }$) guarantees a high
energy outcoming $\tau$ even if from a thinner Earth crust.

\begin{figure}[htbp]
\begin{center}
\includegraphics[angle=270,scale=0.3]{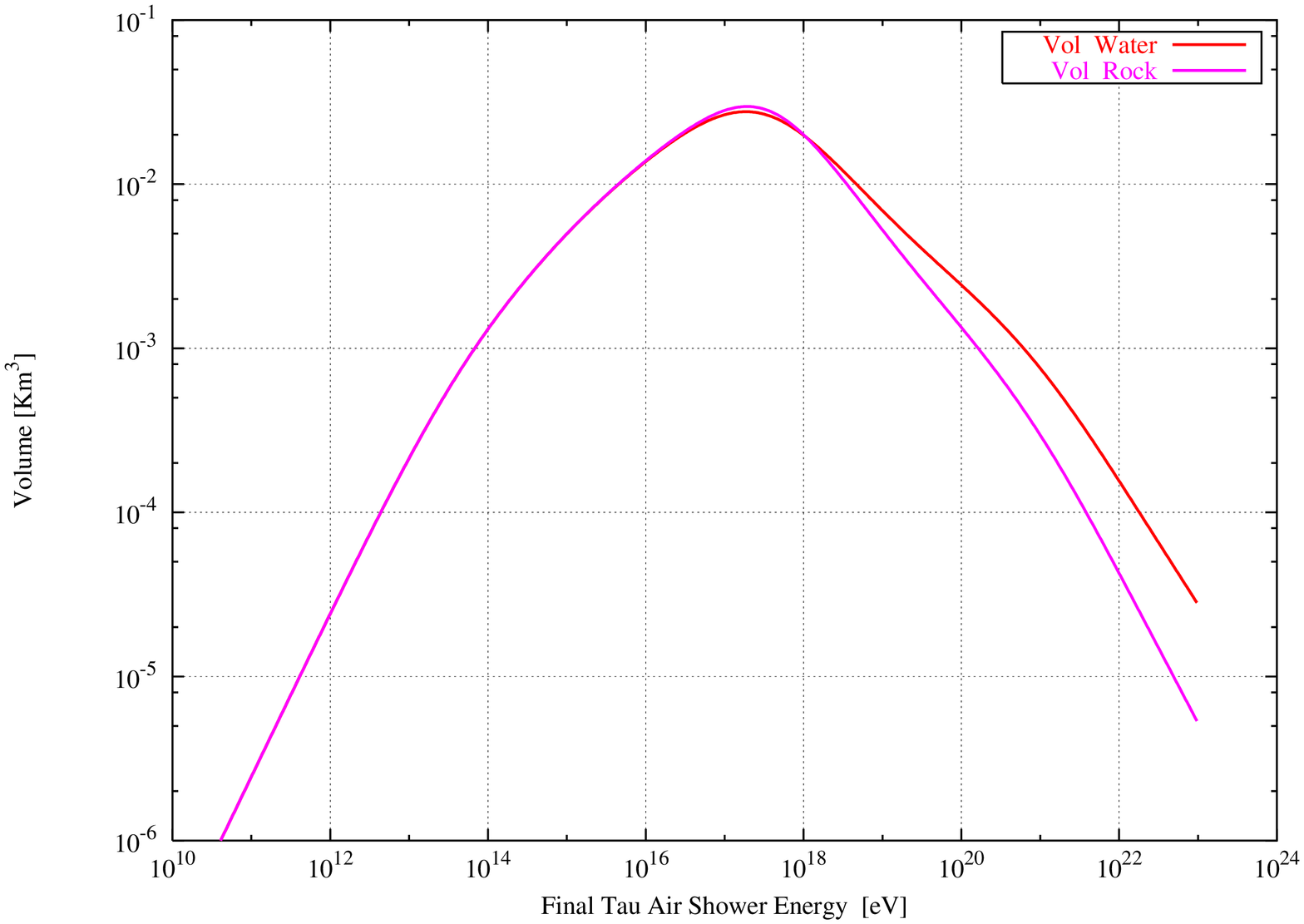} %
\includegraphics[angle=270,scale=0.3]{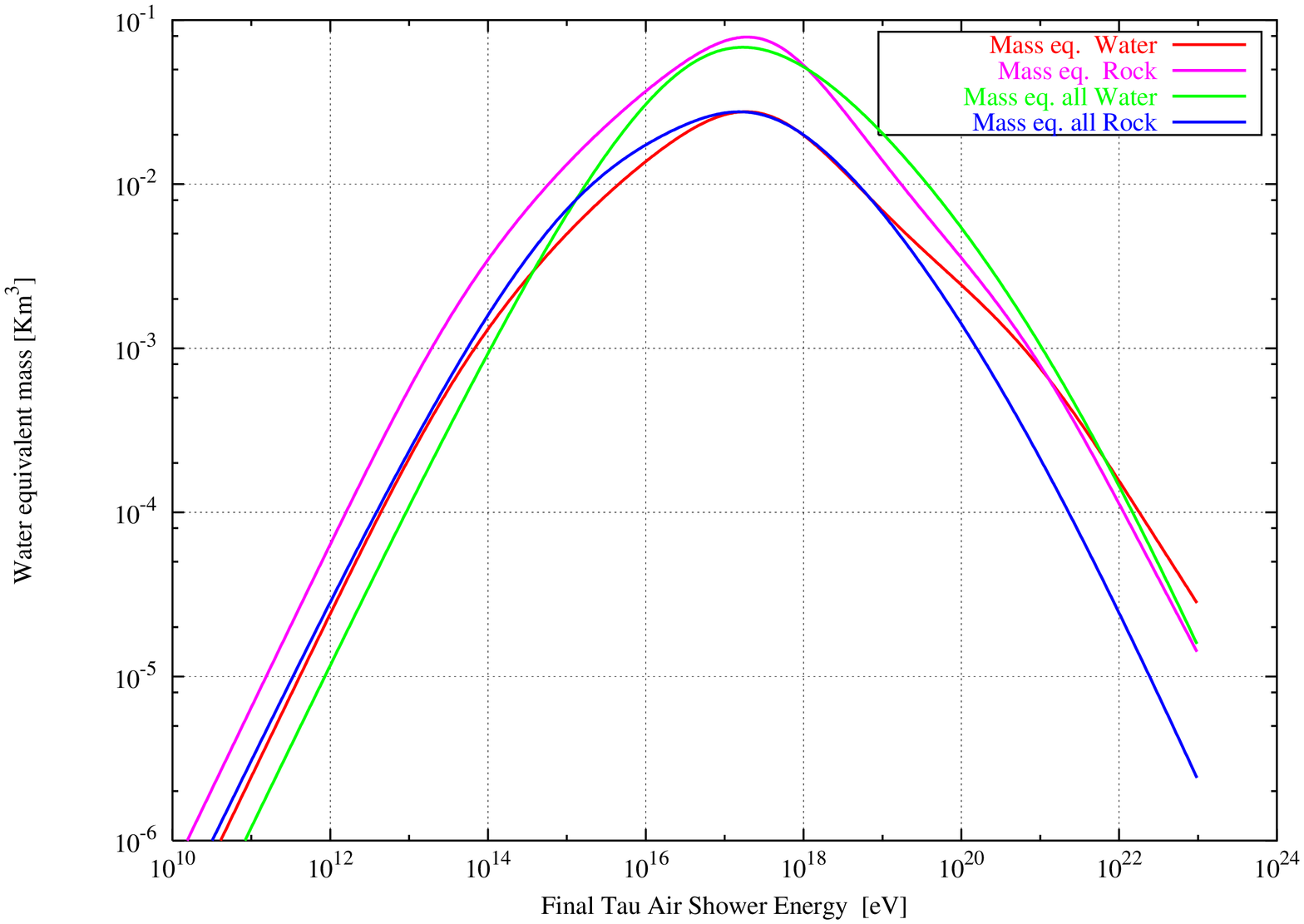}
\end{center}
\caption{\textbf{left)} Effective Volume for UPTAUs and HORTAUs
per km$^2$ unit area neglecting the finite extension of the
horizontal atmospheric layer. Here the calculation has been
performed using the most restrictive interaction length
$L_{\protect\tau (\protect\beta)}$. Therefore on the x axis we
have plotted the final energy of the tau lepton, rather than the
initial neutrino energy as in Fig. \ref{fig7}. Again the two
curves correspond to an outer layer made of water (solid line)
and an outer layer made of
rock (dotted line). \textbf{%
right)} Effective Mass for UPTAUs and HORTAUs per km$^2$ unit area
derived under the same assumptions.  As in Fig. \ref{fig7} the
curves obtained are compared with a simplified model of the Earth,
considered as an homogeneous sphere of water (dashed line) and rock
(big dotted line). Again, note that in these cases the masses have
been obtained under the constraint of a finite extension of  the
terrestrial atmosphere (an horizontal length  of $600$ km).}
\label{fig12}
\end{figure}

The terrestrial cord, $D(\theta ),$ is responsible for the $\nu
_{\tau }$ opacity, and $L_{\nu }$ is the interaction length for
the incoming neutrino in a water equivalent density, where
$L_{\nu_{CC} }=(\sigma_{CC} \,n)^{-1}$. It should be kept in mind
that both $L_{\nu }$ and $D(\theta )$, the water equivalent
cord, depend on the number density $n$ (and the relative matter density $%
\rho _{r}$). We remind that the total neutrino cross section
$\sigma _{\nu }$ consists of two main component, the charged
current and neutral current terms, but the $\tau $ production
depends only on the dominant charged current whose role will
appear later in the event rate number estimate. The interaction
lengths $L_{\tau \beta }$, $L_{\nu_{CC} },$ depends on the
energy, but one should be careful on the energy meaning. Here we
consider an incoming neutrino with energy $E_{\nu _{i}}$, a
prompt $\tau $ with an energy $E_{\tau _{i}}$ at its birth place,
and a final outgoing $\tau $ escaping from the Earth with energy
$E_{\tau _{f}}$, after some energy losses inside the crust. The
final $\tau $ shower energy, which is the only observable
quantity, is nearly
corresponding to the latter value $E_{\tau _{f}}$ because of the negligible $%
\tau $ energy losses in air. However we must be able to infer
$E_{\tau _{i}}$ and the primary neutrino energy, $E_{\nu}$, to
perform our calculation. The effective volume and masses resulting
from Eq. \ref{Veff_lbeta} are displayed in Fig. \ref{fig12}.

\begin{figure}[h]
\begin{center}
\includegraphics[angle=270,scale=0.3]{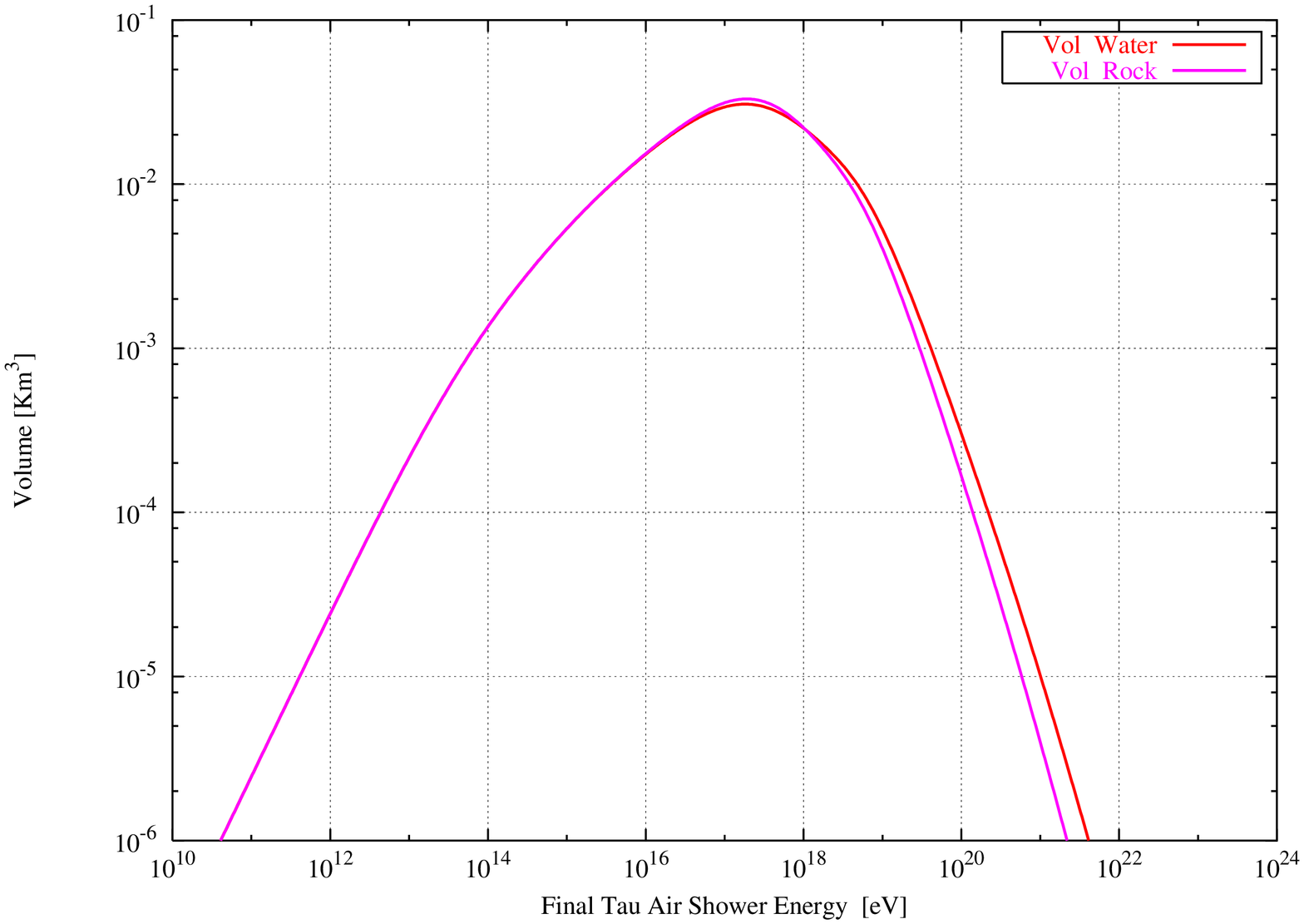} %
\includegraphics[angle=270,scale=0.3]{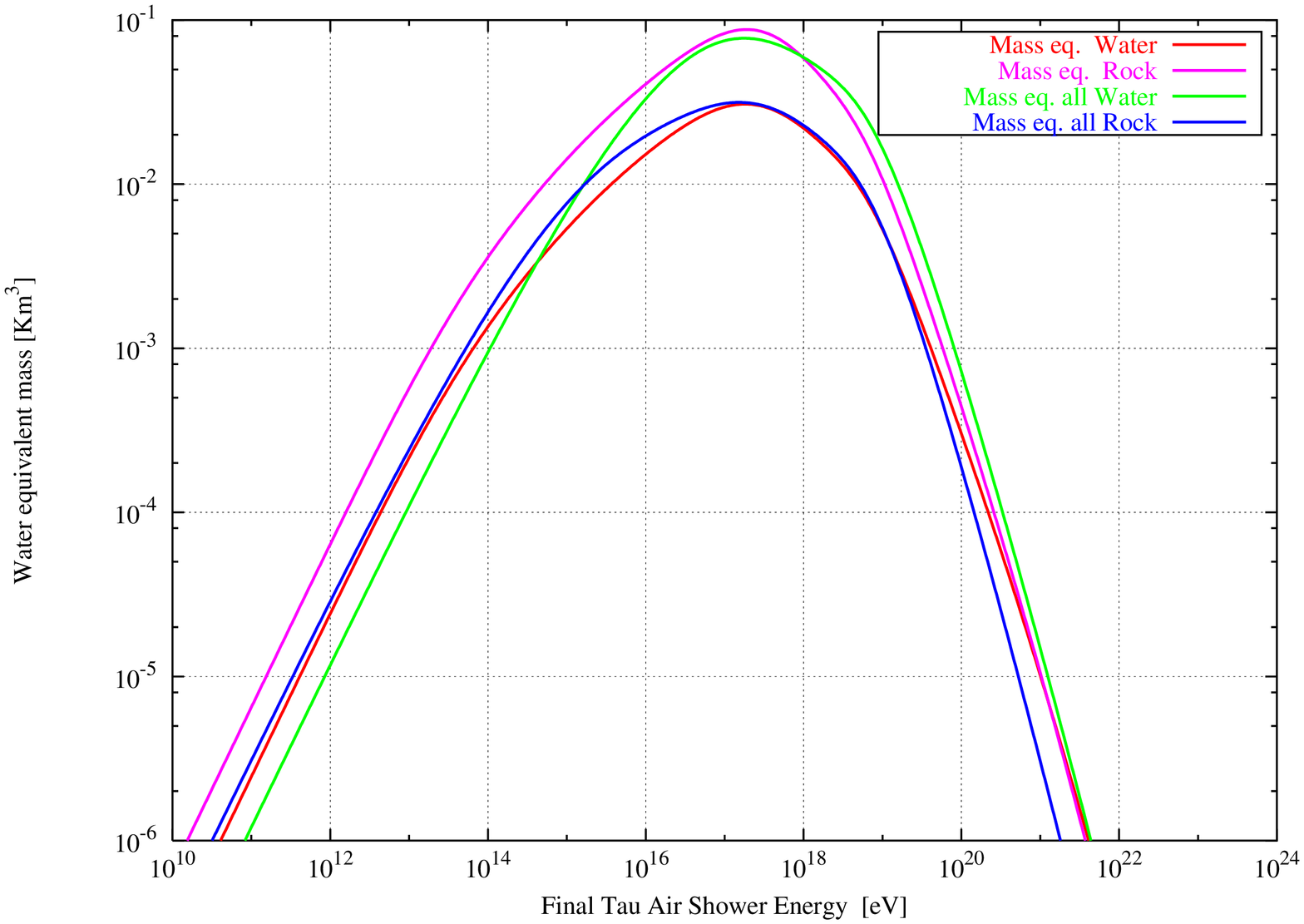}
\end{center}
\caption{\textbf{left)} Effective Volume for UPTAUs and HORTAUs
per km square unit area including the suppression factor due to
the finite extension of the horizontal atmospheric layer. Again,
the calculation has been performed using the most restrictive
interaction length $L_{\protect\tau (\protect\beta)}$ and the
volume is expressed as a function of the final tau energy.
\textbf{right)} The corresponding effective Mass for UPTAUs and
HORTAUs for km square unit area.} \label{fig14}
\end{figure}


However one has to take into account the presence of the Earth
atmosphere which implies a further suppression of the effective
volume especially at high energies. Therefore Eq. \ref{Veff_lbeta}
becomes

\begin{equation}
\frac{V_{Tot}(E_{\tau})}{A}= \left( 1 - e^{- \frac{L_0}{c \tau \gamma_{\tau}}%
} \right) \left(\frac{L_{\tau (\beta)}(E_\tau)}{1-\frac{L_{\tau
(\beta)}(E_{\tau})}{L_{\nu_{CC}}(\eta E_\tau)}}\right)\int^{\frac{\pi}{2}%
}_{0}e^{-\frac{D(\theta)}{L_{\nu_{CC}}(\eta E_\tau)}}\sin{\theta}\cos{\theta}%
d\theta  \label{Veff_lbeta_air}
\end{equation}

where the first term outside the integral guarantees that the relativistic $%
\tau$ length will never exceed the longest inclined path through the Earth's
atmosphere ($L_o\simeq600 \,Km$) corresponding to a characteristic height $%
h_1 \simeq 23 \,Km$ (Fargion 2001b, 2002a).

The effective volume and mass obtained in this case are shown in
Fig. \ref{fig14} for an acceptance of $1$ km$^2$ unit area. Even
if the Effective Volumes and Masses are now reduced these values
still exceeds those of the atmospheric layer above the same $km^2$
area. Moreover it should be emphasized that the detection of
downward neutrino induced air-shower is affected by different
experimental problems. Most of the vertical events  cross a too
small slant depth, they can not develop a full air-shower and
their signature is hidden in the UHECR downward  background.
Therefore HORTAUs are still the most relevant tracers of UHE
$\tau$ neutrinos in the range of energy between $10^{16}$ eV and
$10^{20}$eV.

\begin{figure}[htbp]
\par
\begin{center}
\includegraphics[width=12cm,height=8cm]{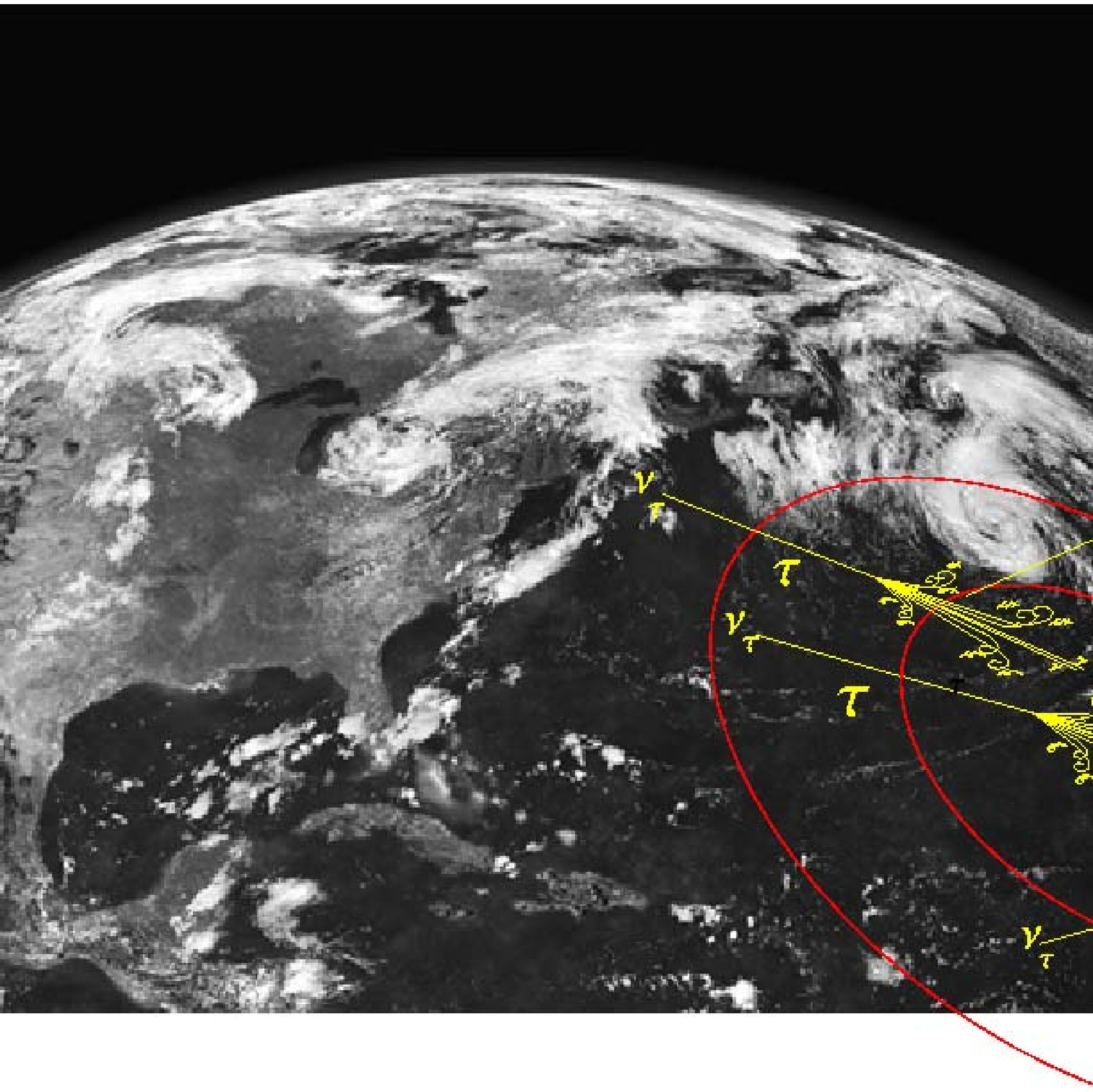}
\end{center}
\caption{A schematic figure showing the field of view (FOV) of
EUSO where $\tau$ air-shower are taking place. We note that the
EUSO threshold energy above $ \sim 10^{19}$ eV implies $\tau$
tracks longer than $500$ km, consequently only UHE neutrino
interacting within a much wider terrestrial crust ($700-900$ km
radius rings) might produce  HORTAUs  observable within the
telescope's FOV. The correspondence between the outcoming $\tau$
along the wide ring (producing the HORTAUs pointing to the EUSO
FOV) and consequent HORTAUs observed within the same FOV,
guarantees the equivalence between the effective volumes defined
by these two different regions. There are three main groups of
HORTAUs: the showers totally included in the FOV, and those that
are partially contained in the FOV (both incoming and outgoing);
the latter will exceed the fully contained ones. The HORTAUs
should extend for a few hundred km in horizontal and reach tens
of kilometers in altitude. Such showers should be opened by
geo-magnetic forces into thin, fan-shaped, charged and neutral
jet tails, mainly observable as forked  and diffused Cherenkov
lights. Their characteristic opening is ruled by the local
geo-magnetic field strength and directionality.}
\label{figEUSOring}
\end{figure}

Among current and future neutrino detectors we would like to
narrow our analysis down to two different projects capable of
indirectly searching Tau Air-Showers: the ground-based arrays of
scintillator/photomultipliers which constitutes the Auger project,
and the space observatory EUSO, whose characteristics have been
described in section \S 3. EUSO may  detect HORTAUs (Horizontal
Tau Air-Showers) originated within a very wide terrestrial skin
volume around its field of view (FOV), corresponding to a few
hundred kilometer radius wider area. A schematic picture its
field of view
 and the kind of air-showers it may be able to observe is shown in
Fig. \ref{figEUSOring}.


\begin{figure}[htbp]
\begin{center}
\includegraphics[angle=270,scale=0.3]{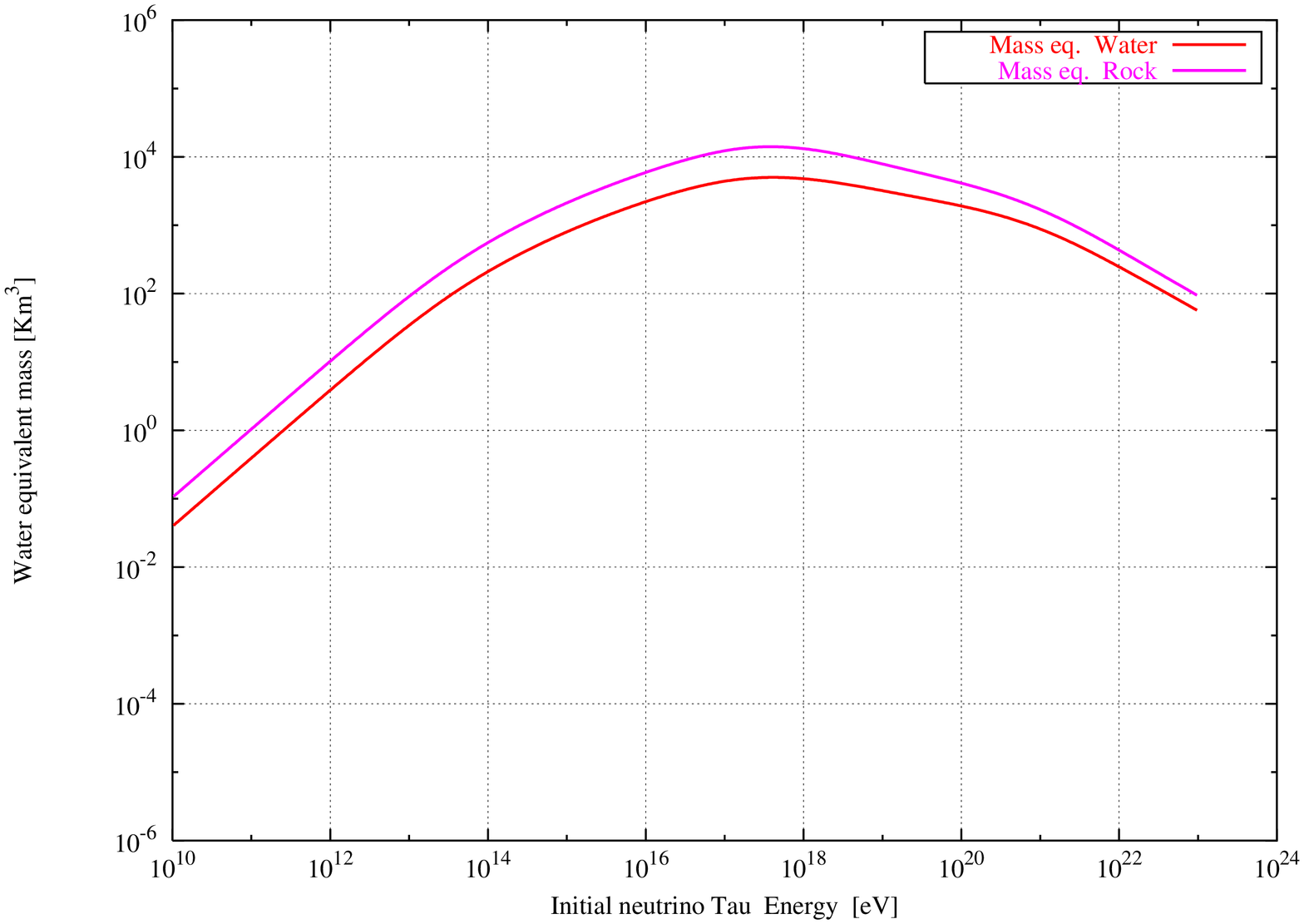}
\includegraphics[angle=270,scale=0.3]{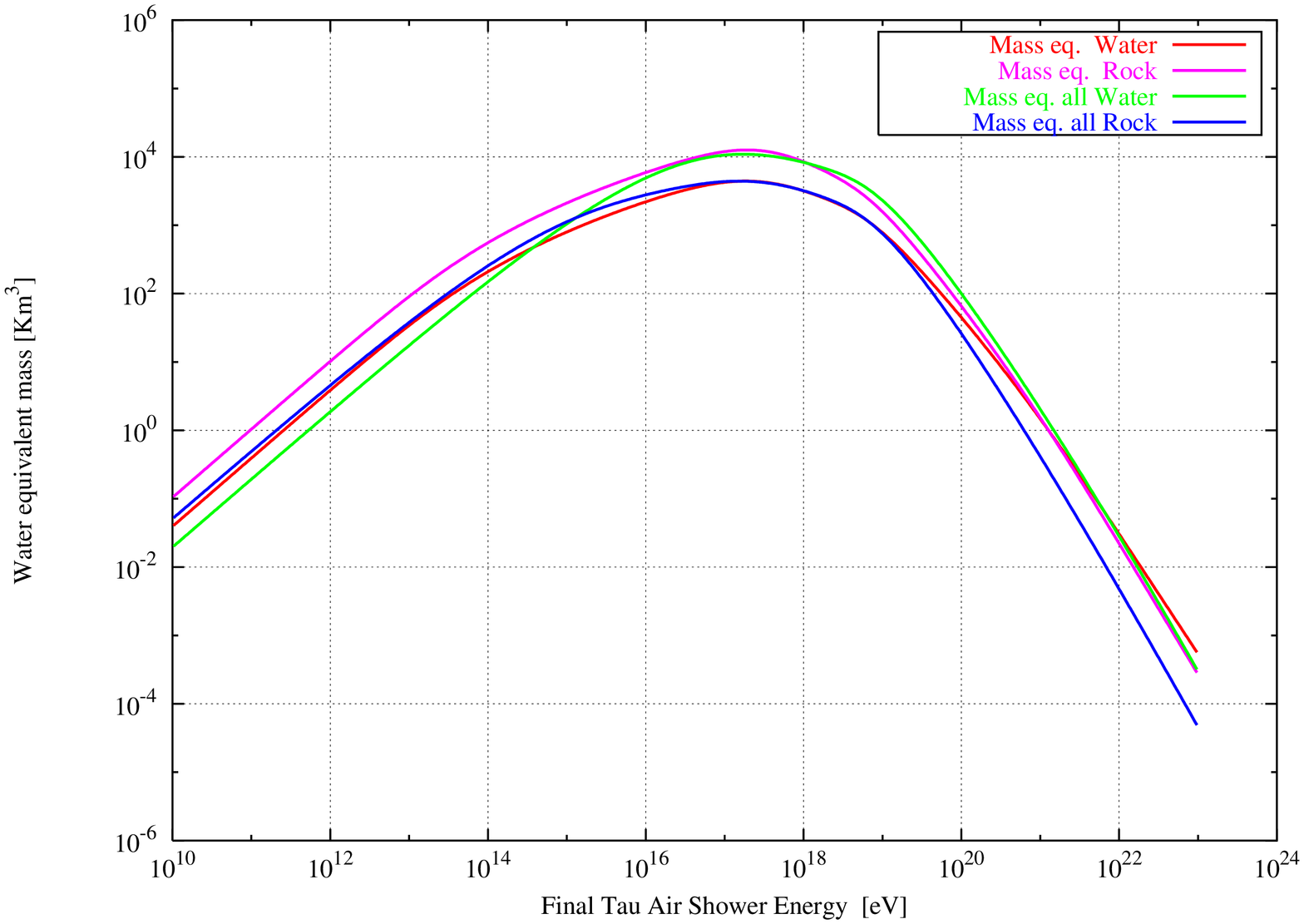}
\end{center}
\caption{{\bf left)} Effective Mass for UPTAUs and HORTAUs for EUSO
as a function of the incoming neutrino energy, calculated with the
interaction length $l_{\tau}$. Here we have neglected the finite
horizontal extension of Earth's atmosphere. The water mass-volume
at $E_{\nu} = 10^{19}$ eV, is $3.2 \cdot 10^3$ $km^3$, while for
the rock the mass-volume is $7.9 \cdot 10^3$ $km^3$. {\bf right)}
Effective Mass for UPTAUs and HORTAUs for EUSO with
$L_{\protect\tau (\protect\beta)}$, considering the finite
horizontal extension of Earth's atmosphere. The water
mass-volume is $7.8 \cdot 10^2$ $km^3$, while for the rock the volume is $%
5.9 \cdot 10^2$ $km^3$. The corresponding mass is nearly $1.6 \cdot
10^3$ water equivalent $km^3$.} \label{fig16}
\end{figure}


After having derived the effective volume and mass for a 1 km$^2$
unit area detector, we finally  show in Fig. \ref{fig16} the
Effective mass for EUSO. Note that the earliest prediction for a
simplified Earth model considered as an homogeneous sphere of
water and rock in the energy range $10^{15} - 10^{19}$ eV are not
too different from our last result (Fargion 2002b), even though as
it appears from the figure, the role of rock and water has now
been inverted. Such a result has not been noticed before (Fargion
2002b, 2002c; BG03). The remarkable huge mass (either for water
and rock) $7.8 \cdot 10^2$ $km^3$ and $1.6 \cdot 10^3$ water
equivalent $km^3$  makes EUSO (even with $10 \%$ duty cycle) the
widest neutrino telescope in Neutrino Astronomy. Other concurrent
experiment (SALSA, ANITA and FORTE) are mainly neutrino
collectors unable to follow their exact arrival direction.

\section{Event Rate for GZK neutrinos with EUSO}

The consequent event rate for incoming neutrino fluxes may be  easily
derived by:

\begin{equation}
\frac{dN_{ev}}{d\Omega dt}= \left( \int \frac{dN_{\nu}}{dE_{\nu}
d\Omega dA dt } \sigma_{N \nu} (E) dE \right) n \rho_r V_{Tot}
\end{equation}

If the cross section is slowly varying with the energy (as it is in the $\nu
N$ reaction), the integral reduces to

\begin{equation}
\frac{dN_{ev}}{d\Omega  dt }= \frac{dN_{\nu} E_{\nu}}{dE_{\nu}
d\Omega dA dt } \sigma_{N \nu} n \rho_r V_{Tot} =
\frac{\phi_{\nu}}{L_{\nu_{CC}}} \rho_{r} V_{Tot}
\end{equation}

\bigskip where $(\sigma_{N \nu} n)^{-1} = L_{\nu CC}$ and   $\phi _{\nu }=\frac{dN_{\nu}E_{\nu}}{dE_{\nu}d\Omega dA dt }
$, and $\rho_r$ is the density of the layer relative to the water.
As it has been mentioned in the previous section we have neglected
the $ \nu _{\tau}- N$ scattering via the neutral current channel.
This process causes a $\nu _{\tau }$ regeneration through a
marginal energy degradation by a factor $(1-y)\simeq 0.8$ where $y$
is the inelasticity parameter (Gandhi et al. 1995, 1998). Although
this term may be responsible for a sequence of $\nu _{\tau }$
regeneration, its contribution to the number of events
 is mainly taken into account by the use of the charged current cross section
(and not by the additional neutral current contribute).


\begin{figure}[h]
\begin{center}
\includegraphics[angle=0,scale=0.6]{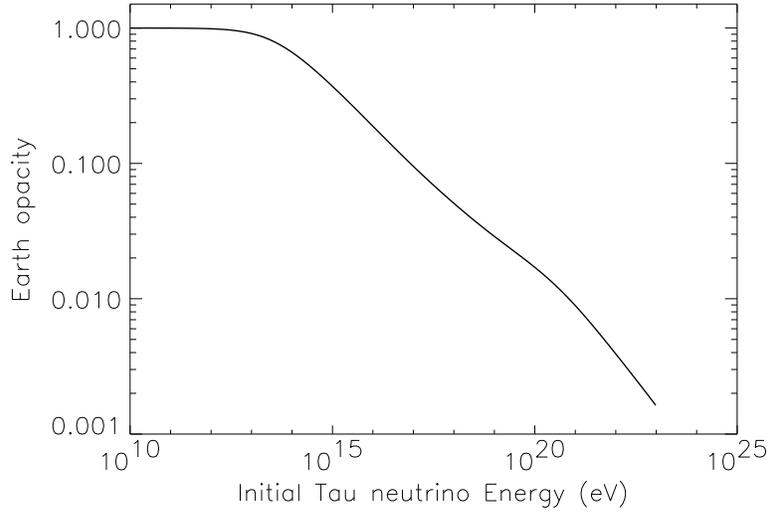}
\caption{The Earth opacity to upward neutrinos as a function of the
initial $\nu_{\tau}$ energy.} \label{opacita}
\end{center}
\end{figure}


After having introduced the effective volume we can estimate the outcoming
event number rate for EUSO for any given neutrino flux. In particular this
general expression will be displayed assuming a minimum GZK neutrino flux $%
\phi _{\nu }\simeq \phi _{UHECR}\simeq 5\cdot
10^{-18}cm^{-2}s^{-1}sr^{-1}$ comparable to the observed UHECR
one at the same energy ($E_{\nu }=E_{UHECR}\simeq 10^{19}eV$).
The assumption on the flux may be changed at will  and the event
number will scale linearly according to the model.

However, the initial incoming $\nu$ flux $\phi _{\nu
}=\frac{dN_{\nu}E_{\nu}}{dE_{\nu}d\Omega dA dt }$ is suppressed
when neutrinos cross the Earth by a shadow factor

\[ S = \int_0^{\pi/2} e^{- \frac{D(\theta)}{L_{\nu CC}}} cos\theta d\theta \]

shown in Fig. \ref{opacita}. This function defines the ratio
between the final outgoing upward neutrinos, $\phi_{\nu_f}$ and
the initial downward $\phi_{\nu_i}$, $S(E_{\nu_i}) =
\phi_{\nu_f}/ \phi_{\nu_i}$.


\begin{figure}[p]
\includegraphics[angle=0,scale=0.7]{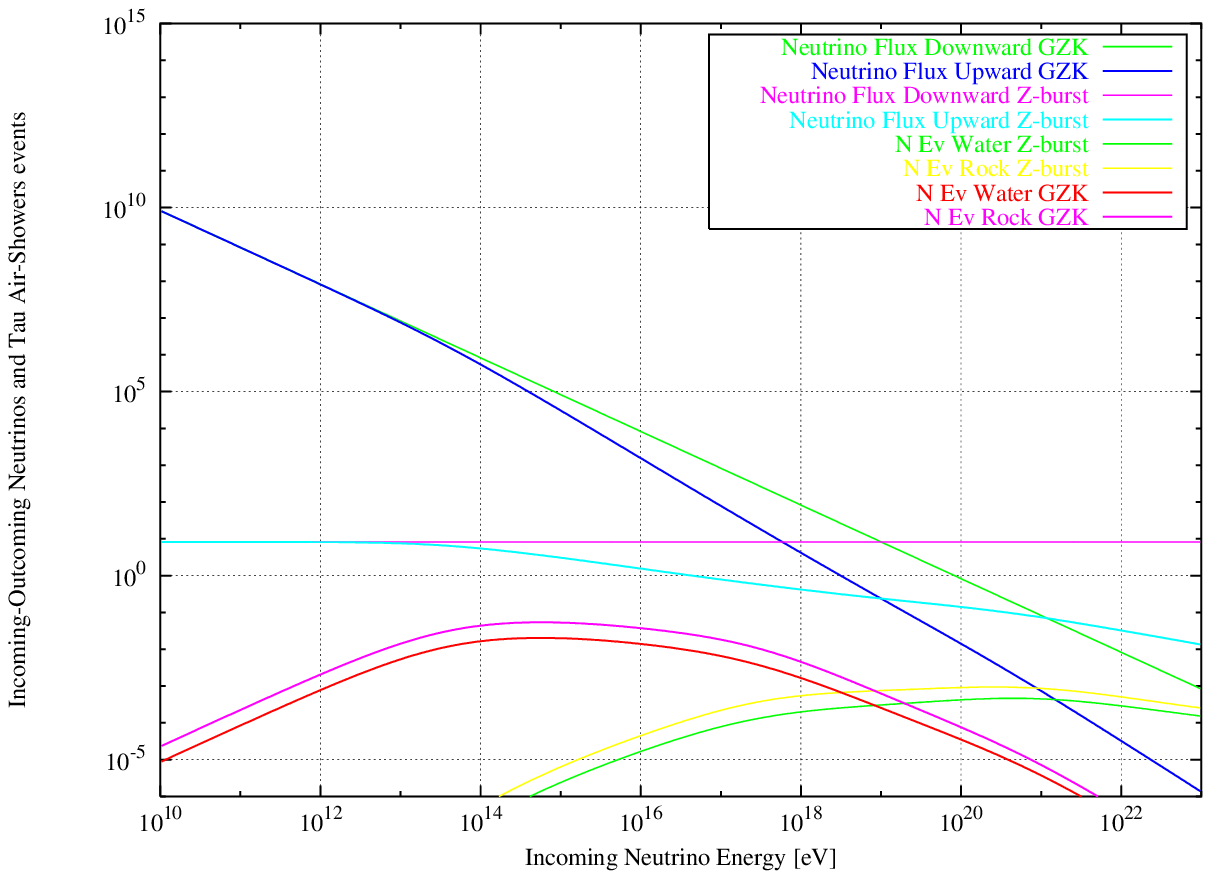}
\includegraphics[angle=0,scale=0.7]{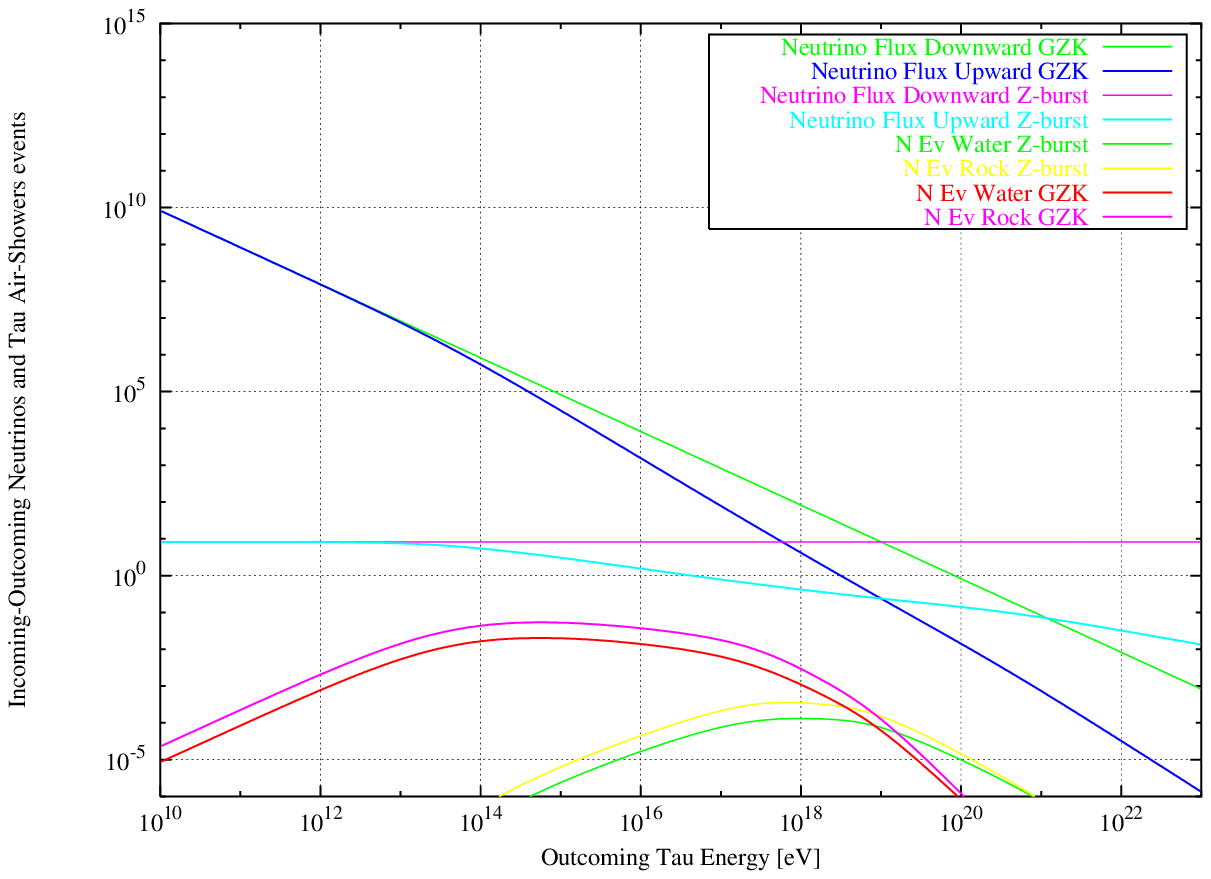}
\caption{ {\bf left)} Number of HORTAUs Events per km$^2$ and for
one year data collection  as a function of the incoming neutrino
energy excluding the finite extension of the horizontal
atmospheric layer, with a $\tau$ lepton interaction length given
by $l_{\tau}$ (see Fig. \ref{l_tau}).  Here GZK and Z burst
neutrino fluxes refers to the incoming upward and downward
neutrino spectrum $\propto E^{-2}$ (inclined dashed line), and
$\propto E^{-1}$ respectively (horizontal dotted-line). The
outcoming neutrino curves are suppressed  above energies $10^{14}$
eV compared to the incoming $\nu$ fluxes by the shadow factor
defined in Fig. \ref{opacita}. For each flux the number of events
is derived for an outer layer of both
rock and water. At $10^{19}$ eV we find $N_{ev}^{water}= 2.55 \cdot 10^{-4}$ ($\phi_{\protect\nu} E_{\nu} / 50$ eV cm$^{-2}$ s$%
^{-1}$ sr$^{-1}$) and $N_{ev}^{rock}= 6.26 \cdot 10^{-4}$ ($\phi_{\protect\nu} E_{\nu} / 50$ eV cm$^{-2}$ s$%
^{-1}$ sr$^{-1}$). {\bf right)} Number of events as a function of
the outgoing lepton tau for $L_{\tau (\beta)}$, including the
finite extension of the horizontal atmospheric layer. Again the
results are compared to a GZK/Z-burst neutrino flux, and they are
displayed for an outer layer made of rock and
water. In this case $N_{ev}^{water}= 0.62 \cdot 10^{-4}$ ($\phi_{\protect\nu} E_{\nu} / 50$ eV cm$^{-2}$ s$%
^{-1}$ sr$^{-1}$) and $N_{ev}^{rock}= 1.25 \cdot 10^{-4}$ ($\phi_{\protect\nu} E_{\nu} / 50$ eV cm$^{-2}$ s$%
^{-1}$ sr$^{-1}$). } \label{figNevkm2}
\end{figure}

In Fig. \ref{figNevkm2} we show the expected number of event per
km$^2$ to more easily compare our results with those of other
authors. The results have been obtained for two different
scenarios. On the left-hand side we display the number of events
having neglected the presence of the atmospheric layer, with the
choice of $l_{\tau}$ as interaction length. Under this assumption
we calculate the number of events as a function of the energy of
the incoming neutrino. On the right-hand side we have included the
Earth's atmosphere and we have used $L_{\tau (\beta)}$, thus we
may express the results as a function of the final $\tau$ lepton
energy. Both scenarios have been calculated with two incoming
neutrino fluxes, whose power law spectrum is proportional to
$E^{-2}$ (which approximates at highest energies the GZK flux) and
$E^{-1}$ (that  mimics the Z-burst flux) respectively. In general
one may imagine a power law $E^{-\alpha}$ , $0<\alpha <1$ within
the two extreme laws above.  These incoming neutrino fluxes are
also shown in Fig. \ref{figNevkm2}, as well as the final neutrino
fluxes $\phi_{\nu_f} = S(E_{\nu_i}) \phi_{\nu_i}$ for both
models. One has to bear in mind that the final number of events
for UPTAUs-HORTAUs is determined by the $V_{eff}$, $M_{eff}$
functions (Eqs. 1 - 4) which already take into account the Earth
opacity to neutrinos with the $\nu$ interaction length $L_{\nu
CC}$. Therefore $N_{ev} \propto \phi_{\nu_i}$  rather than
$N_{ev} \propto \phi_{\nu_f}$.

 At $E^{19}$ eV in the more restrictive
approximation we obtain the following number of events for an
area $km^2$ and in one year data collection :


\[  N_{ev}^{water}= 2.55 \cdot 10^{-4} (\phi_{\protect\nu} E_{\nu}
/ 50 eV cm^{-2} s ^{-1} sr^{-1}) \: \: \: \: \: N_{ev}^{rock}=
6.26 \cdot 10^{-4} (\phi_{\protect\nu} E_{\nu} / 50 eV \, cm^{-2}
\, s ^{-1} \, sr^{-1})    \]

\begin{figure}[htbp]
\begin{center}
\includegraphics[angle=270,scale=0.3]{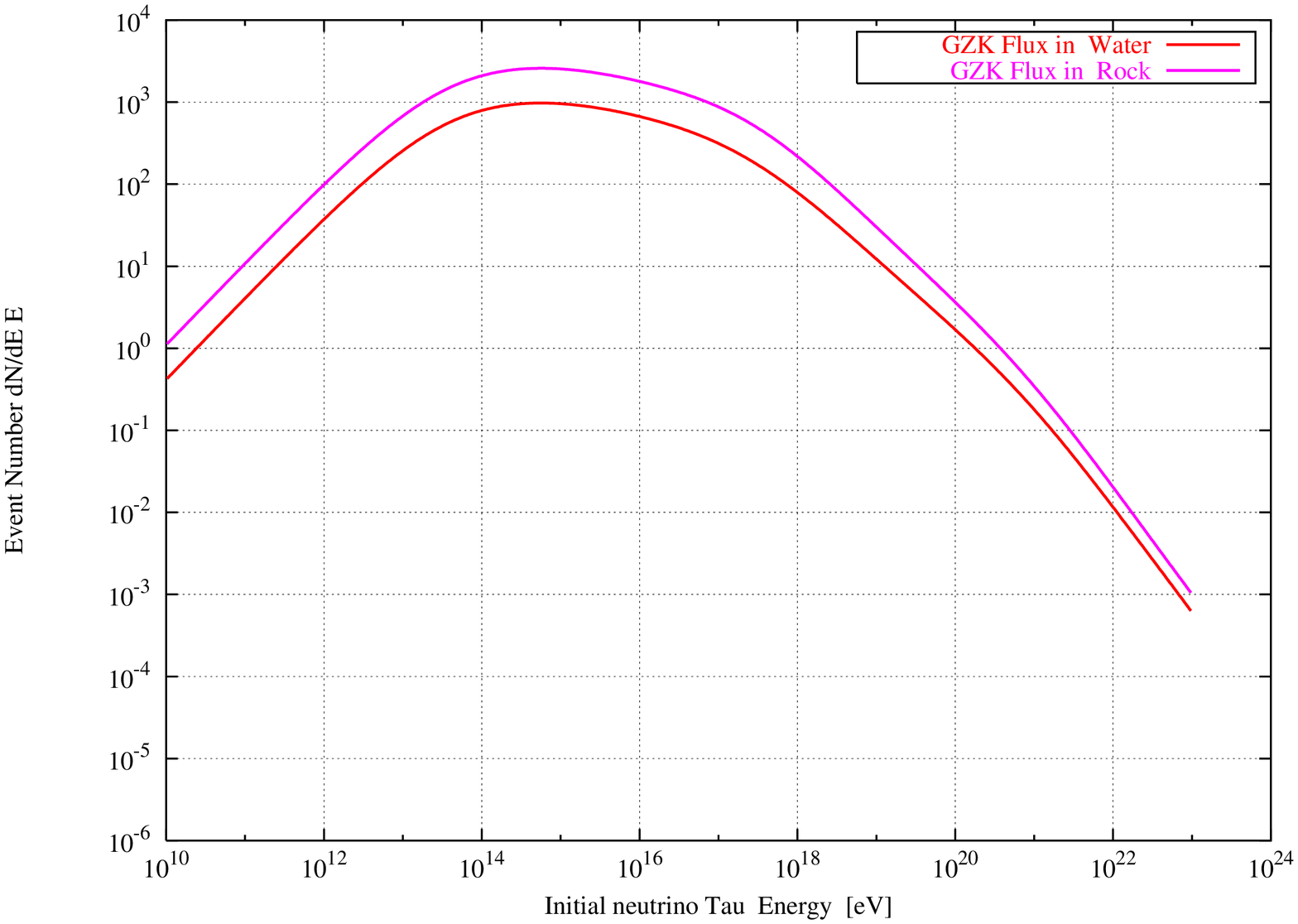}
\includegraphics[angle=270,scale=0.3]{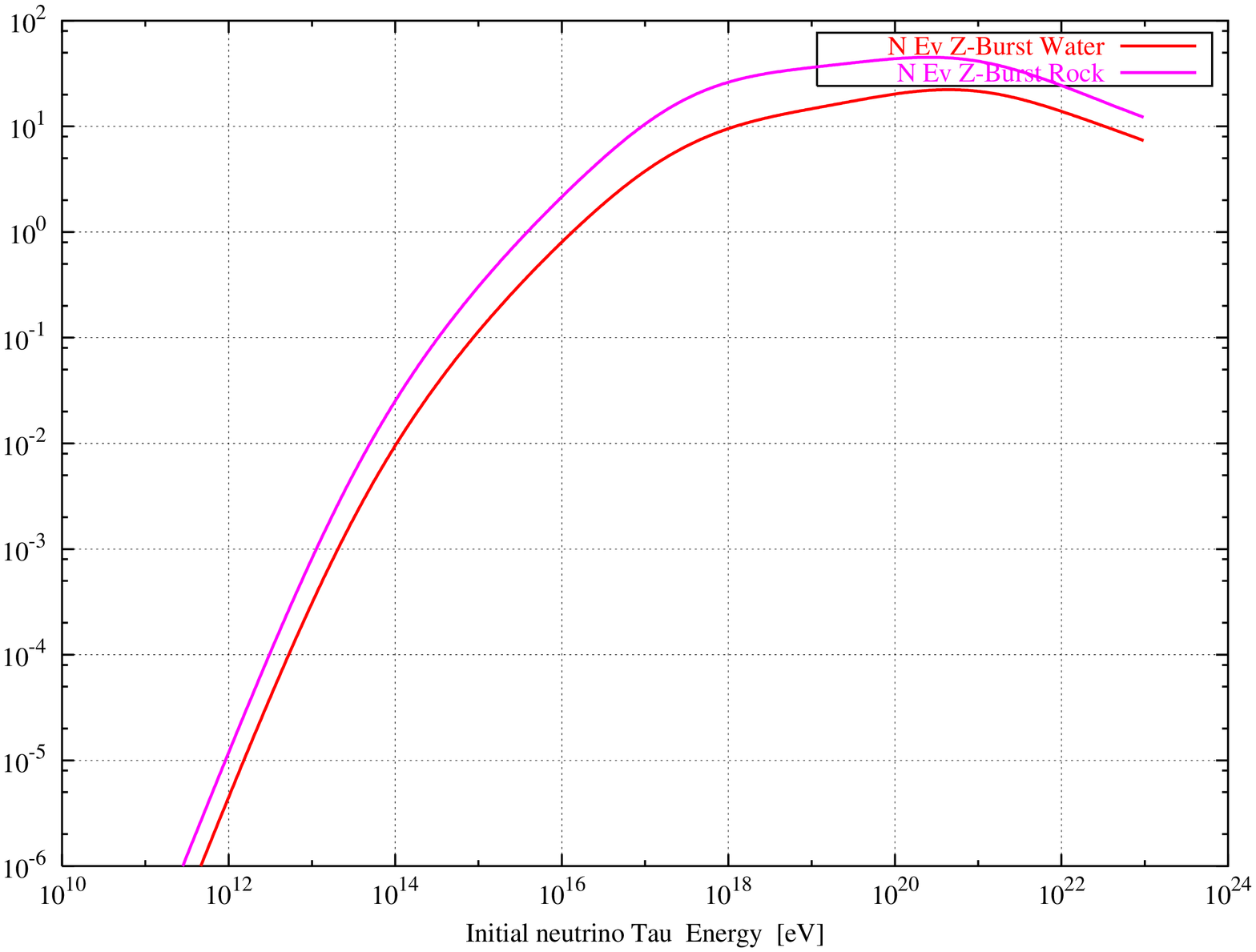}
\end{center}
\caption{Number of EUSO Event for HORTAUs in 3 years record as a
function of the incoming neutrino tau energy, having neglected the
finite extension of the horizontal atmospheric layer, with
$l_{\tau}$ as interaction length. As above, we have used both a
GZK
 {\bf (left)} and a Z-burst {\bf (right)}  neutrino flux. At $E_{\protect\nu} = 10^{19}$ eV,
the expected event number is $12.2$ ($\protect\phi_{\protect\nu}
 E_{\nu} / 50$ eV cm$^{-2}$ s$^{-1}$ sr$^{-1}$) for the water
and 30 ($\protect\phi_{\protect\nu} E_{\nu} / 50$ eV cm$^{-2}$
s$^{-1}$ sr$^{-1}$) for the rock.} \label{fig18}
\end{figure}

As regards EUSO, the general expected event rate around the same
energy range is given by:

\begin{eqnarray}
N_{ev}\,= 5\cdot 10^{-18}cm^{-2}s^{-1}sr^{-1}\,
\left({\frac{V_{eff} \rho_r}{L_{\nu CC}}}\right) (2
\pi\,\eta_{Euso} \Delta t) \left( \frac{\Phi_{\nu} E_{\nu}}{50 eV
cm^{-2} s^{-1} sr^{-1}} \right) \left( \frac{\eta
E_{\tau}}{10^{19} \,  eV} \right)^{- \alpha} %
\end{eqnarray}

where $\eta_{Euso}$ is the duty cycle fraction of EUSO, $\eta_{Euso} \simeq
10\%$, $\Delta \,t\ \simeq 3$ $years$.
The consequent number of events  for its area
($1.6\cdot10^{5}Km^2$) is displayed in Fig. \ref {fig18}
assuming both a flat power law ($\phi_{\nu} \propto\,
E_{\nu}^{-2}$) and a harder Z-Burst spectra $\phi_{\nu}
\propto\,E_{\nu}^{-1}$. These spectra are very simple and nearly
model independent. They also fit a Berezinsky (1990) $\phi_{\nu}
\propto\, E_{\nu}^{-2}$ or a Waxman-Bachall (1997, hereafter WB97)
spectra for GRB  as well as the GZK flux. Here we have neglected
the presence of the atmosphere above the portion of the Earth
surface covered by EUSO. The event number at $E_{\nu} = 10^{19}$ eV
in such an approximation is for the EUSO area and for three years
data collection:

\[
N_{ev}^{water} = 12.2 (\phi_{\nu} E_{\nu} / 50 \; eV \, cm^{-2} \,
s^{-1} \, sr^{-1}) \: \: \: \: \: \: \: \: N_{ev}^{rock} = 30
(\phi_{\nu} E_{\nu}/50 \; eV \, cm^{-2} \, s^{-1} \, sr^{-1})
\]

 Such number of events greatly exceed previous results by at least two orders of magnitude (Bottai et al. 2003)
 but are below more recent estimates (Yoshida et al. 2004).

\begin{figure}[htbp]
\includegraphics[angle=270,scale=0.3]{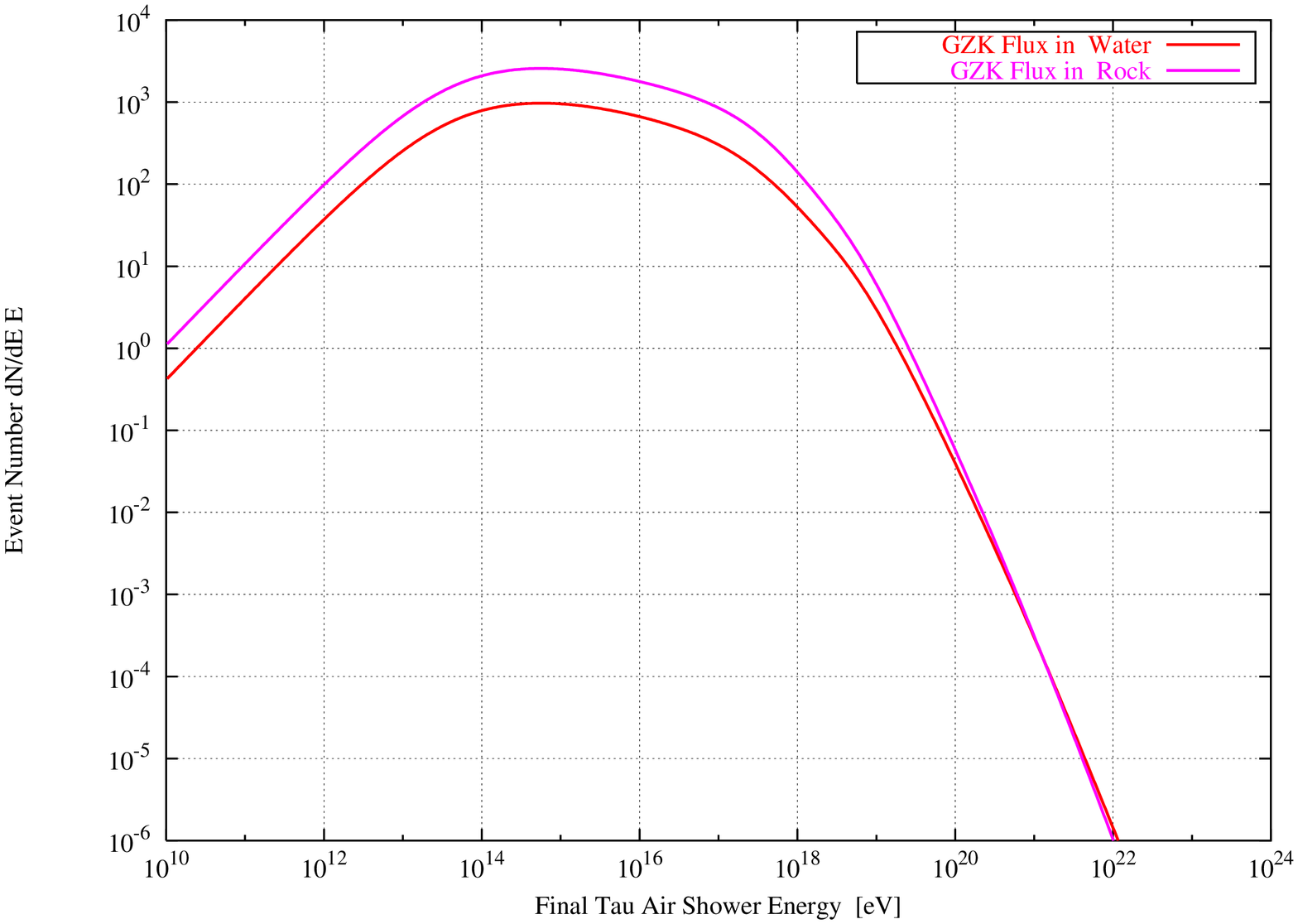}
\includegraphics[angle=270,scale=0.3]{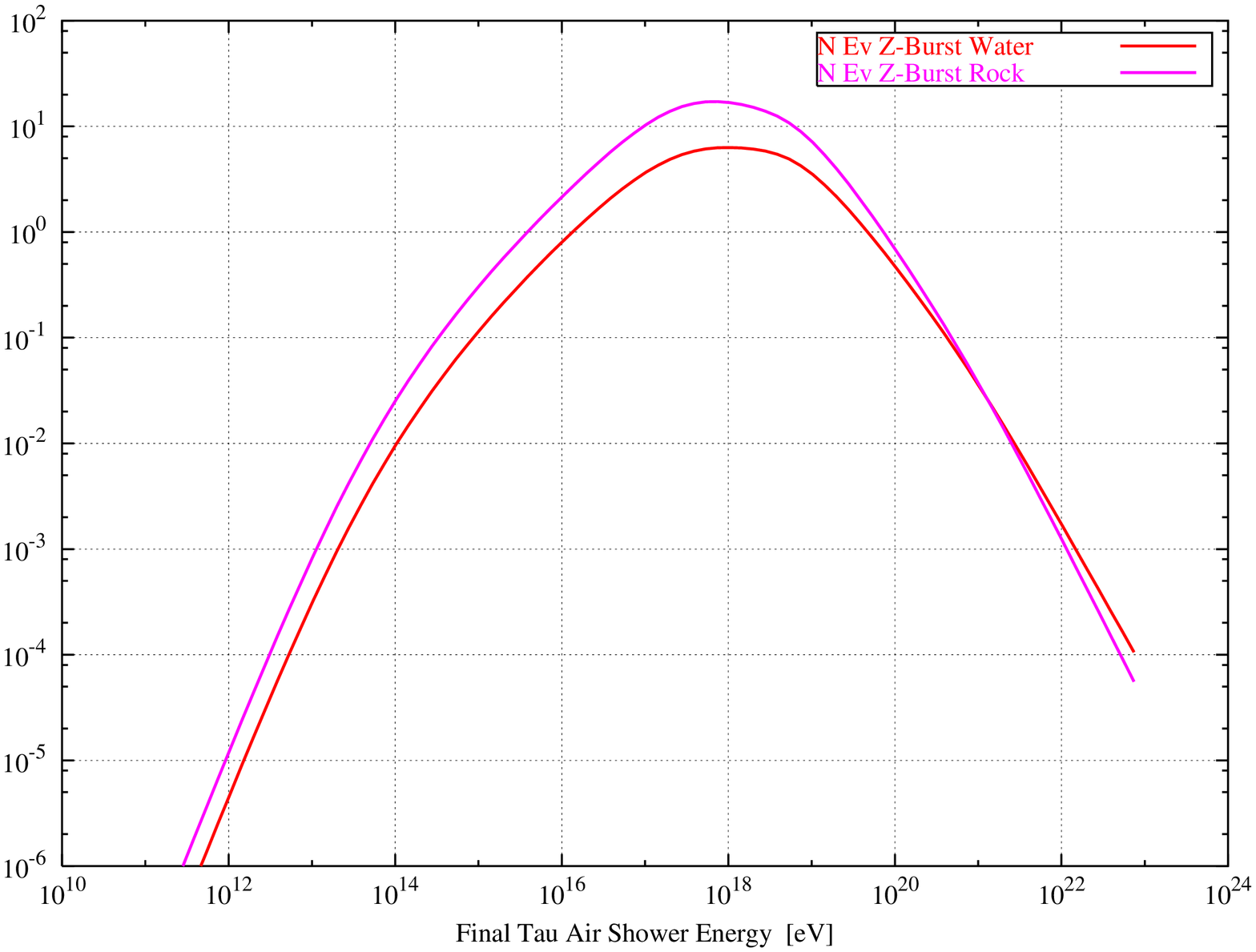}
\caption{Number of EUSO Event for HORTAUs in 3 years record as a
function of the outgoing lepton tau ($L_{\tau (\beta)}$ as
interaction length), including the finite extension of the
horizontal atmospheric layer. At energy $E_{\tau} = 10^{19}$ eV,
the event number is $N_{ev}= 3.0$ ($\protect\phi_{\protect\nu}
E_{\nu} / 50$ eV cm$^{-2}$ s$^{-1}$ sr$^{-1}$) for the
water and $N_{ev}= 6.0$ ($\phi_{\protect\nu} E_{\nu} / 50$ eV cm$^{-2}$ s$%
^{-1}$ sr$^{-1}$) for the rock. Again, we show the resulting
number of events for two different neutrino fluxes: $\propto
E^{-2}$ {GZK \bf (left)} and $\propto E^{-1}$  (Z-burst {\bf
right)}.} \label{fig19}
\end{figure}

When the Earth's atmosphere is included in the calculation of the
effective volume (Eq. \ref{Veff_lbeta_air}), this causes a
general suppression of the event rate, especially at $E_{\nu} >
1.2 \cdot 10^{19}$ eV. At such energies the $\tau$ decay length
exceeds the maximal thickness of the atmospheric layer ($\simeq
600$ km), reducing the possibility of $\tau$ shower detection
with EUSO. The expected number of event in such a case is shown
in  Fig. \ref{fig19} for both a GZK and Z-burst fluxes. As one
can see from Fig. \ref{fig19} we obtain

\[
N_{ev}^{water} = 3.0 (\phi_{\nu} E_{\nu} / 50 \; eV \, cm^{-2} \,
s^{-1} \, sr^{-1}) \: \: \: \: \: \: \: \: N_{ev}^{rock} = 6.0
(\phi_{\nu} E_{\nu} / 50 \; eV \, cm^{-2} \, s^{-1} \, sr^{-1})
\]

which is yet larger than the unity for the three years scheduled
for the EUSO project. Because most of the HORTAUs would be
observed partially contained (inward or outward the EUSO Field of
view ) the total number of events might be  doubled.

It should be kept in mind that the plots of the EUSO number of
events in Fig. \ref{fig18}, \ref{fig19} are already suppressed by
a factor $0.1$ due to the minimal duty cycle factor $\eta_{EUSO}$.

\begin{figure}[htbp]
\begin{center}
\includegraphics[angle=0,scale=0.6]{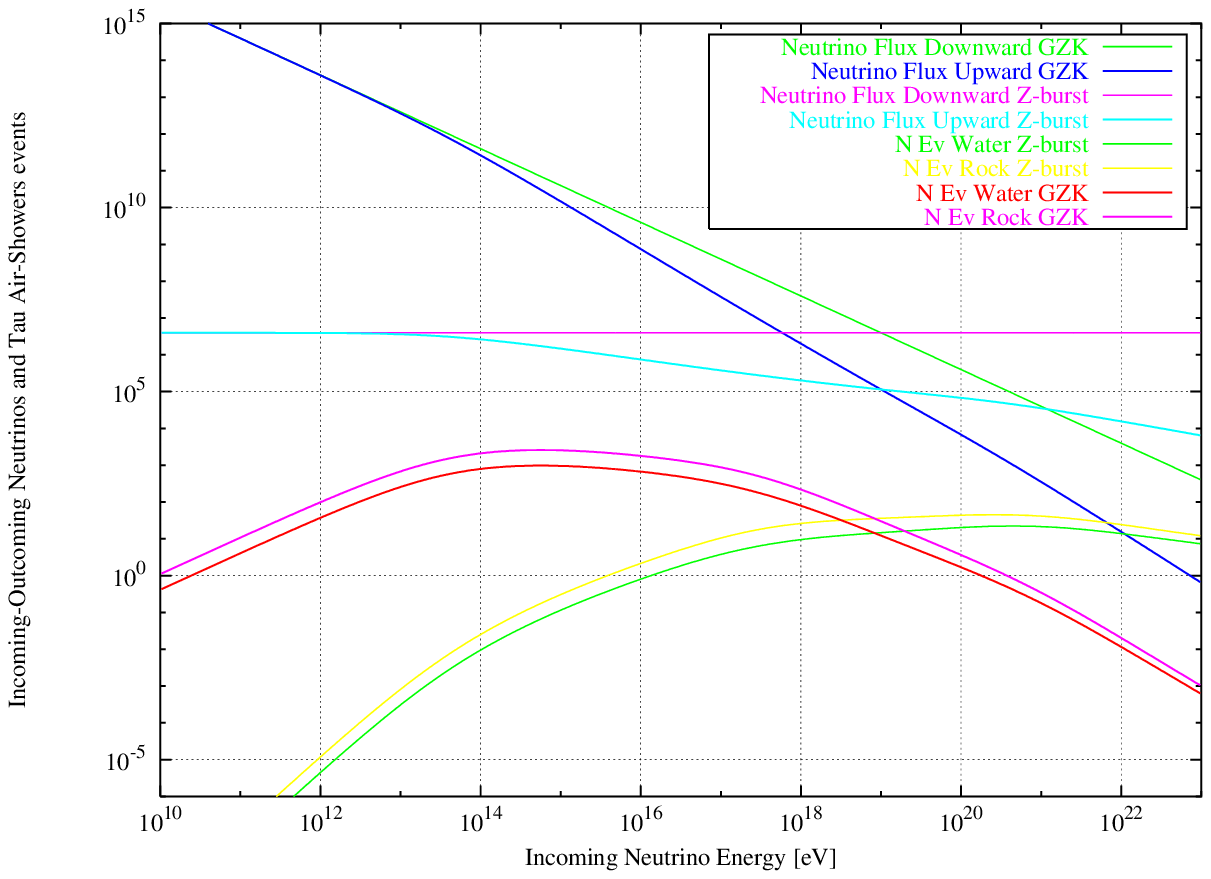}
\includegraphics[angle=0,scale=0.6]{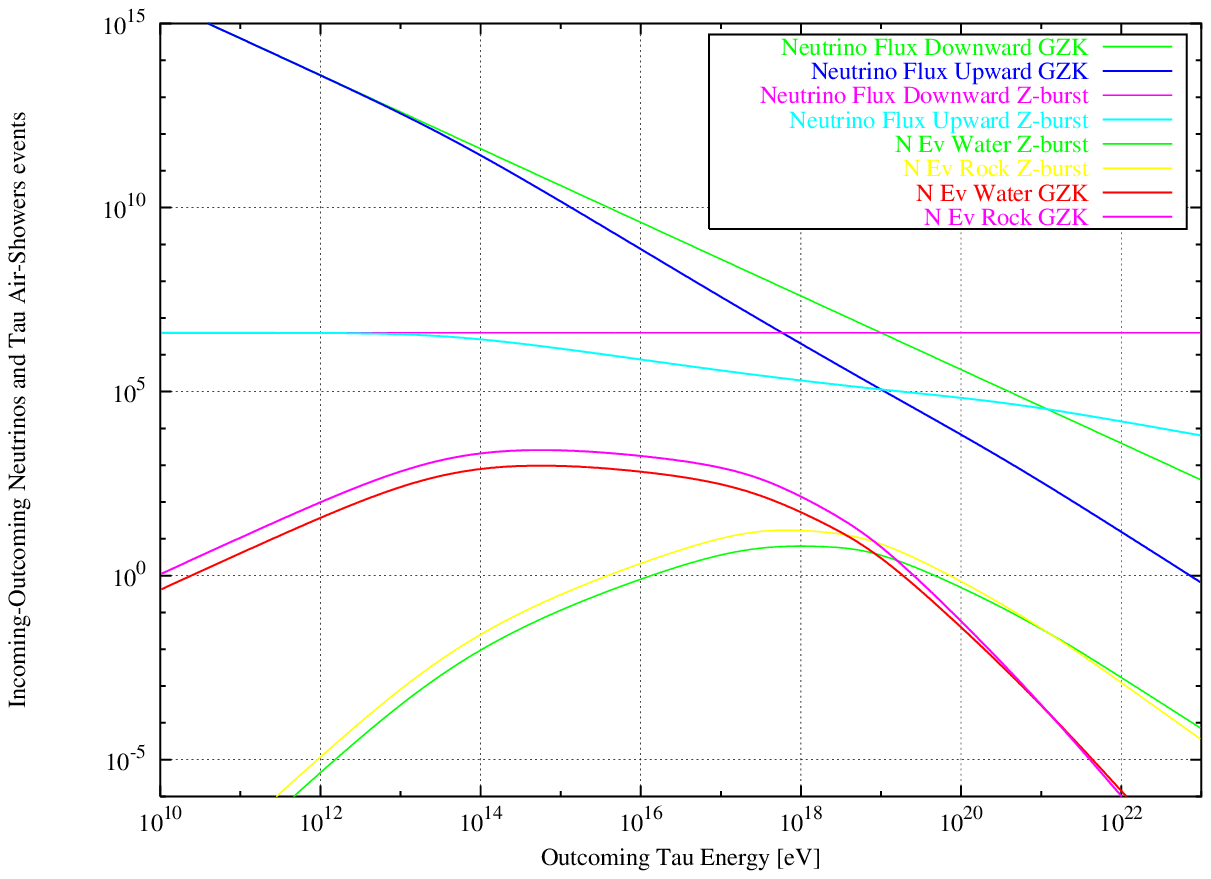}
\end{center}
\caption{Number of EUSO Event for HORTAUs in 3 years record
compared to the upward and downward neutrino flux (GZK, mostly
comparable to a flux $\propto$ $E^{-2}$, and Z-burst $\propto
E^{-1}$).  The curves describing the number of events on the left
and righthand side are the same as Fig. \ref{fig18} (using
$l_{\tau}$ and neglecting the atmosphere) and Fig. \ref{fig19}
(using $L_{\tau (\beta)}$ and including the atmosphere)
respectively.} \label{fig20}
\end{figure}

In Fig. \ref{fig20} we summarize our results with and without the presence
of the Earth's atmosphere and we compare the obtained number of events with
the upward and downward neutrino fluxes for a GZK and Z-burst spectrum.

It is worth noticing that the number of horizontal $\tau $ showers
observable by EUSO in three years record time is above unity at
energies below $E_{\tau}\simeq 10^{19}eV$, with a peak around
$10^{18}$ eV. This is quite a remarkable result which may
strengthen the case of extending the EUSO energy sensitivity below
$10^{19}$ eV by enlarging the size of the telescope. The rich
Cherenkov diffused lights from HORTAUs might produce enough photons
to allow a lower (than $10^{19}$ eV ) energy threshold.

\begin{figure}[htbp]
\begin{center}
\includegraphics[angle=0,scale=0.8]{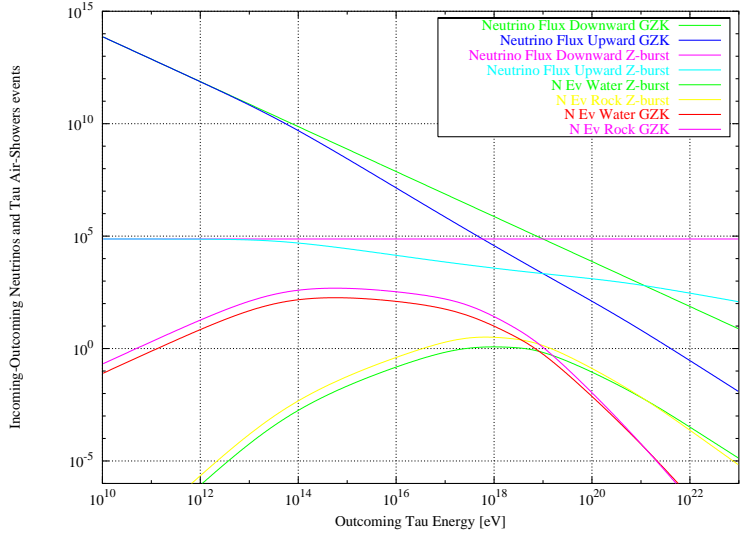}
\end{center}
\caption{Number of Events of HORTAUs expected to be detected by
Auger in 3 years record, having neglected the finite extension of
the horizontal atmospheric layer. Again, we show the results for
two different neutrino fluxes: GZK and Z-burst. As one can see the
number of events is slightly above the unity at $E = 10^{19}$ eV.
However at energy as low as $E = 10^{18}$ eV in GZK model the
number of events increases to $26.3$ for a rock layer: the most
remarkable signature will be a strong azimuthal asymmetry
(East-West) toward the high Andes mountain chain. The Andes shield
UHECR (toward West) suppressing their horizontal flux. These
horizontal showers originated in the atmosphere at hundreds of km,
are muon-rich as well as poor in electron pair and gammas and are
characterized by a short time of arrival. On the other hand the
presence of the  mountain chains at $50-100$ km from the Auger
detector will enhance by a numerical factor $2-6$ (for each given
geographical configuration) the event number respect to the East
direction. These rare nearby (few or tens of km) tau air-shower
will be originated by Horizontal UHE neutrinos interacting inside
the Andes. These showers will have a large electron pair, and gamma
component greater than the muonic one, and a dilution in the time
of arrival.}\label{fig21}
\end{figure}

\begin{figure}[htbp]
\begin{center}
\includegraphics[angle=0,scale=0.8]{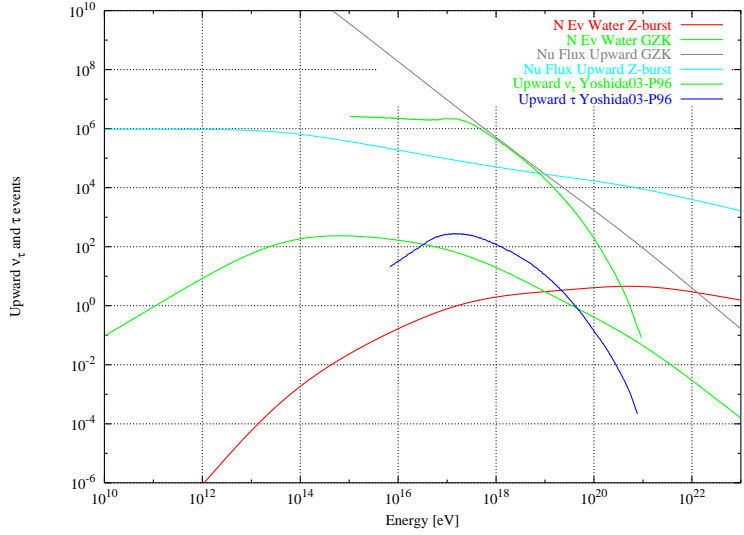}
\end{center}
\caption{Number of EUSO Event for HORTAUs in 3 years record,
having neglected the finite extension of the horizontal
atmospheric layer, compared to the results of Yoshida et al.
(2004). Here the assumed a lower GZK and Z-Burst like  spectra ($\phi_{\protect\nu} E_{\nu} / 10$ eV cm$^{-2}$ s$%
^{-1}$ sr$^{-1}$) and we are evaluating the events assuming a
water target, while Yoshida considered an almost comparable
($0.9$ lower density) ice mass. There is a partial disagreement:
Yoshida predictions are at least a factor $2.7$ at $E = 10^{19}$
eV and nearly an order of magnitude at  $E = 10^{18}$ eV larger
than ours. At PeVs-EeVs energies where comparison with Gandhi et
al. (1998) was possible we found a general agreement. On the
contrary  the Bottai \& Giurgula. (2003) (see their Fig.11,c)
expectations are in complete disagreement with us, leading to a
number of events at least two order of magnitude smaller than
ours.} \label{fig22}
\end{figure}



\begin{table}[t]
\begin{tabular}{|c|c|c|c|c|}
  \hline
   & $N_{ev}^{water} \:$ & $N_{ev}^{rock} \:$  & $N_{ev}^{water} \:$  & $N_{ev}^{rock} \:$  \\
   &  atmosphere included & atmosphere included & no atmosphere & no atmosphere \\
  \hline
 km$^2$ (1 yr)$^{-1}$ & 0.62$ \cdot 10^{-4}$ & $1.25 \cdot 10^{-4}$    & $2.55 \cdot 10^{-4}$  & $6.26 \cdot 10^{-4}$   \\
  Auger (3 yr)$^{-1}$ & 0.56 & 1.13 &  2.3 & 5.64 \\
  EUSO (3 yr)$^{-1}$  & 3.0 & 6.0  & 12.2 & 30.0 \\
 \hline
\end{tabular}
\caption{Number of events at $10^{19}$ eV obtained for different
detectors and periods of data collection. In the first
   two columns  the calculation includes the finite extension of the atmospheric layer
  where the HORTAUs shower may take place  and we adopted the most restrictive $L_{\tau (\beta)}$ as the interaction length,
  while in the latter two columns we have neglected the presence of the atmosphere and we have used $l_{\tau}$.
  The assumed incoming neutrino fluence within the GZK or Z-burst model at energy $10^{19}$ eV is   $\phi_{\nu} \simeq 50 \cdot E^{-1}$ eV $cm^{-2}$  $s^{-1}$
 $sr^{-1}$  }
\end{table}

\section{Conclusions}

Horizontal and Upward $\tau $ neutrinos emerging from the Earth and
the consequent detection of $\tau $ showers in air may represent an
alternative way to investigate the neutrino astrophysics  and
possibly to open a Neutrino Astronomy at energies greater than
$10^{15}$ eV. The conversion of $\tau $ neutrinos into $\tau $
leptons coming out from a mountain chain (Fargion, Aiello, \&
Conversano 1999) or from the Earth (Fargion 2002a; Feng et al.
2002; Bertou et al. 2002) may lead to electromagnetic and hadronic
showers producing an amplified signal of gamma,X,muon bundles,
electron pairs and Cherenkov or fluorescent photons. Contrary to
any downward $\nu$ induced air-shower the detection of upward tau
air-showers is not constrained by the presence of a  high
background noise. We present an exact analytical procedure to
calculate the expected rate of up-going tau air-showers applied to
the characteristics of large area detectors such as Auger and EUSO.
We have introduced an effective volume and mass, which are
independent of the neutrino fluxes, describing the Earth volume
where neutrino-nucleon ($\nu _{\tau }-N$) interactions may produce
emerging $\tau $ leptons. We have calculated such effective volume
and mass taking into account the exact terrestrial structure, as a
function  of both the initial $\nu _{\tau }$ and  the final $\tau $
energy (the latter being the only observable quantity). To obtain a
realistic prediction of effective target volumes and masses  we
considered the $\tau $ energy losses while propagating within  the
Earth and the atmosphere finite size where $\tau$ might decay in
flight. We prove the dominant role of HORTAUs and UPTAUs volume and
mass respect to those of atmospheric layers in the very relevant
range $10^{16}-10^{19}$ eV.  Secondly, we show for the first time
that when we consider a multi-layer structure of the Earth as
predicted by Dziewonski (1989), the equivalent mass for an outer
layer made of rock is dominant compared to the water, contrary to
simplified all-rock/all-water Earth models (Figs. \ref{fig12},
\ref{fig14}, \ref{fig16}) and  previous studies (Bottai \& Giurgula
2003).

The EUSO space Observatory  designed to detect  UHECR may  also
discover  HORTAUS and UPTAUS. We find that the effective
HORTAUs-UPTAUs  mass for water and rock in EUSO at $10^{19}$ eV
is $780$ $km^3$ and $1600$ $km^3$ water equivalent (and, at
$10^{18}$ eV,  $3.18 \cdot 10^3$ $km^3$, $8.48 \cdot 10^3$ $km^3$
respectively). The average Earth mass observed by EUSO is $1.02
\cdot 10^3$ $km^3$. Even considering the EUSO duty cycle
efficiency as low as $0.1$ these huge masses ($\simeq 100 km^3$)
are making the EUSO telescope the widest neutrino future
detectors at HORTAUs highest energies.

We show that  neutrino signals may be well detected by EUSO,
obtaining an event number rate at $E_{\tau} = 10^{19}$ eV
 which is  at least $N_{ev}^{water}$ = 3.0 ($\phi_{\nu} E_{\nu} / 50$ eV cm%
$^{-2}$ s$^{-1}$ sr$^{-1}$)  in the most conservative scenario
assuming that the outer layer of the Earth is made of water
($\rho_r = 1.02$). Moreover we find that $N_{ev}^{rock}$ = 6.0
($\phi_{\nu} E_{\nu} / 50$ eV cm$^{-2}$ s$^{-1}$ sr$^{-1}$),  for
an outer layer relative density $\rho_r = 2.65$. The consequent
average number of events in three years is $<N_{ev}>$ = 4.0
($\phi_{\nu} E_{\nu} / 50$ eV cm$^{-2}$ s$^{-1}$ sr$^{-1}$),
obtained with a minimal WB97 or GZK neutrino flux $\propto E^{-2}$.
If one considers the HORTAUs events partially contained in the  FOV
of EUSO (see Fig. \ref{figEUSOring}) they must be at least doubled.
 A lower energy threshold $E_{\tau } \simeq 10^{18}$ eV in GZK models
  possible to achieve for Cherenkov diffused tracks, may lead to a significant increase of the event number up
to an order of magnitude. Geo-magnetic bending of HORTAUs and
UPTAUS at high quota will generally produce a fan-shaped shower "
polarized " along the plane orthogonal to the direction of the
local magnetic field. Such a signature, due to the magnetic
splitting of the different components of the shower ($e^{\pm}$,
$\mu^{\pm}$, $\gamma$) may be detected by EUSO and its
polarization axis is an additional criterion to distinguish such
events. The more abundant HIAS (High Altitude Air-Showers), see
Fig. \ref{fig3}, event rate even nearly one-two order of magnitude
above HORTAUs rate, can not hide the HORTAUs events because of
their characteristic downward signature (Fargion, Khlopov et al.
2003, Fig.32 ).

 However  UPTAUs (at PeV energies) arise at a negligible rate (even for an observing period
of  three years) because of the very narrow beam $\Delta\Omega
\simeq 2.5 \cdot 10^{-5}$ $rad$ needed to point upward towards the
EUSO telescope. Therefore the expected number of events is $N_{eV}
\simeq N_{Uptaus} \cdot \Delta \Omega / \Omega = 0.025 $ for the
water and $N_{eV} \simeq N_{Uptaus} \cdot \Delta \Omega / \Omega =
0.08 $ for the rock.

 We also consider the possibility of detection of HORTAUS and UPTAUS
with Auger, and according to our calculation we obtain with no duty
cycle cut-off a number of events which is at $E_{\tau } \simeq
10^{19}$ eV only slightly above the unity (see Fig. \ref{fig21} and
Table 1).

We find that our results differ from those of  very recent studies
(Yoshida et al. 2004, Bottai \& Giurgola 2002). As regards Yoshida
et al. (2004), shown in Fig. \ref{fig22}, we have obtained a
number of events that is at least 3 times lower even in our less
restrictive scenario, where the atmospheric layer has not been
considered, and where we have used longest $l_{\tau}$ as the
interaction length. At energy band  $10^{16}-10^{18}$ eV   the
discrepancy might be due to the fact that Yoshida et al. (2004)
consider as a detection all those events where tau leptons
propagate through the $km^3$ volume. In our analysis instead, we
consider only the $\tau$'s that are produced inside the thin layer
(mainly defined by $l_{\tau}$) just below the $km^2$ and that are
able to emerge from such a surface area. When we introduce the
more realistic and restrictive  $L_{\tau (\beta)}$ interaction
length and we take into account the presence of the atmosphere,
our rate of events (now as a function of $E_{\tau_f}$ ) at
$E_{\tau} \sim 10^{19}$ eV (see Table 1 and Fig. \ref{fig22}) is
about 12 times lower than Yoshida and collaborators' one. At
$E_{\tau} \sim 10^{18}$ there is still a difference of about one
order of magnitude between our (lower) and their (larger) event
rate.

On the contrary, the comparison of our results with BG03 ones
leads to a larger discrepancy in the opposite way. We infer that a
standard rock outer layer provides a higher number (by a factor
2.6) of events compared to a layer made of water, while according
to BG03 the efficiency of the two matter densities is inverted
(by a factor larger than 2). Secondly the expected number of their
events appears to be much more suppressed, by about two orders of
magnitude, for both the water and  the rock, assuming a GZK or a
WB97 neutrino flux. The direct comparison between the results of
BG03 and Yoshida and collaborators implies an even larger gap.

Finally we emphasize the evidence of an expected high number of
events at energies $E_{\tau} \lesssim 10^{19}$ eV. Therefore the
most relevant Horizontal Tau Air-Shower, HORTAUs at GZK energies
will be better searched and revealed at a lower threshold. These
events may originate within a huge ring around the EUSO FOV whose
surface is $A \geq 6.6 \cdot 10^6 km^2$. The horizontal $\tau$'s
decay occur far away from such a ring  ($\geq 550$ km), inside the
FOV of EUSO, and at high altitudes ($\geq 20-40$ km), and they
will give signatures clearly distinguishable from any other
downward horizontal UHECR. Therefore we suggest ($a$) to improve
the fast pattern recognition of Horizontal Shower Tracks to
discriminate HIAS as well as  HORTAU showers appearing as a
sequence of dots with a forked signature; ($b$) to enlarge the
Telescope Radius to reach energy thresholds lower than $10^{19}$
eV where HORTAU neutrino signals are enhanced, mostly in the
optical wavelengths where Cherenkov photons are produced. ($c$)
To improve the angular resolution within an accuracy
$\Delta\theta \leq 0.2^o$, and the error on the measurement of
the shower altitude to $\Delta h \leq 2$, km to better
disentangle HIAS from HORTAUs. ($d$) to  search for the HORTAUs
enhancement along the highest density geological sites such as
the highest volcanos, or mountain chains. Because of the better
transparency of the water to $\nu_{\tau}$  compared to the rock,
such HORTAUs enhancement would be in principle more evident for
largest mountain chains  close to the sea.  The most remarkable
sites could be found in North America (Rocky Mountains), central
America (Sierra Madre), but in particular the Andes mountains in
South America, and the Indonesian Peninsula in Asia show the
highest density contrast and they are bounded by massive
submarine depths. These continental shelves must enhance greatly
the HORTAUs rates. If the Auger experiment would improve
horizontal shower resolution at low energy ($\simeq 10^{18}$ eV),
it   may benefit of this natural rock barrier observing a
peculiar East-West asymmetry (Fargion, Aiello, \& Conversano
1999; Fargion 2002a; Bertou et al. 2002) at EeV energy band.

To conclude large effective volumes are necessary for neutrino
detectors in order to reach useful sensitivity, given the
extremely low flux and weak interactions of UHE neutrinos. We have
shown that EUSO is a very promising mission because it allows to
inspect HORTAUS form a large effective volume, of the order of a
thousand (or because duty cycle, at least a hundred) cubic
kilometers water equivalent. Other projects such as SALSA (Gorham
et al. 2002a) and ANITA (Gorham et al. 2002b) have a large
effective telescope area, corresponding to neutrino huge
detection volumes nearly comparable with EUSO. But they are
mostly neutrino collecting detectors with poor angular
resolution.  EUSO on the other hand can discriminate the
directionality of the neutrino signatures, and as we have
definitively proved, it has the ability to find out at least half
a dozen of events when we assume the most conservative and
guaranteed  GZK neutrino flux.

\section{Appendix}

 The final results we have presented in this paper have been first
tested and constrained with the analysis of simpler approximate
scenarios which we have used to set upper and lower bounds on our
results. First if we consider the Earth as an homogeneous sphere
made of water (or rock) one finds  (Fargion 2002b,  2003):

$$
\frac{V_{eff}}{A_{\oplus}}=
\int_0^{\frac{\pi}{2}}\frac{(2\,\pi\,\,R_{\oplus}\cos\theta)
\,l_{\tau}\,\sin{\theta}}{2\,\pi\,R^2_{\oplus}}\cdot e^{-
\frac{2\,R_{\oplus}\,\sin{\theta}}{L_{\nu_{\tau}}}}\,R_{\oplus}\,d\theta\,=
\left({\frac{L_{\nu_{\tau}}}{2\,R_{\oplus}}}\right)
^2\,l_{\tau}\int_0^{\frac{2\,R_{\oplus}}{L_{\nu_{\tau}}}}t\cdot
e^{-\,t}d\,t
$$


\begin{eqnarray}
V_{eff-Max}= A_{Euso}\,\left({1-e^{-
\frac{L_0}{c\,\tau_{\tau}\,\gamma_{\tau}}}}\right)
\,\left({\frac{L_{\nu_{\tau}}}{2\,R_{\oplus}}}\right)^2 \cdot
l_{\tau}\left[{1\,-\,e^{-
\frac{2\,R_{\oplus}}{L_{\nu_{\tau}}}}(1\,+\,\frac{2\,R_{\oplus}}{L_{\nu_{\tau}}})
}\right] \label{V_allwater}
\end{eqnarray}

while if we take into account only a thin outer layer of water
(or rock) not deeper than $4.5 km$, (considering all the inner
terrestrial shells of infinite densities):

\begin{eqnarray}
V_{eff-Min}= \,A_{Euso}\,\left({1-e^{-
\frac{L_0}{c\,\tau_{\tau}\,\gamma_{\tau}}}}\right)
\,\left({\frac{L_{\nu_{\tau}}}{2\,R_{\oplus}}}\right)^2\cdot
l_{\tau}\left[{1\,-\,e^{- \frac{2\,R_{\oplus} sin(\theta_1)
}{L_{\nu_{\tau}}}}(1\,+\,\frac{2\,R_{\oplus}
sin(\theta_1)}{L_{\nu_{\tau}}}) }\right] \label{V1shell}
\end{eqnarray}

Where $\theta_1$ is the angle between the tangential plane to the
surface of the Earth at the horizon and to  the inner layer (ocean
- rock) we are considering, and it is $\theta_1 \cong 1.076 ^o$.



The above geometrical quantities have been already defined in the
text. It is possible to see that the effective volume in the low
energy approximation reduces to:

\begin{eqnarray}
V_{eff}&=&\,A_{Euso}\,\left({1-e^{- \frac{L_0}{c\,\tau_{\tau}\,\gamma_{\tau}}%
}}\right) \,\frac{l_{\tau}}{2}
\end{eqnarray}

 This is the simplest result that may be  derived in a direct
 way, and it
 guarantees a link between the most sophisticated analytical
integral (Eq. \ref{eq_vol_ltau}) and the upper and lower bounds
set by Eqs. \ref{V_allwater} and \ref{V1shell}.

Indeed, to calculate the effective volume in a more precise way
one has to consider the Earth as a sequence of shells. For the
first shell one obtains  the same result as in Eq. \ref{V1shell}.

\[
\frac{V_{1}(E_{\tau})}{A}=\frac{l_{\tau}}{\left(1-\frac{l_{\tau}}{%
L_{\nu_{1_{w}}}}\right)} \left(\frac{L_{\nu_{1_{w}}}}{2R_{\oplus}}%
\right)^{2} \left(1-e^{-\frac{L_0}{l_{\tau_2}}}\right)\left[1-e^{- \frac{%
2R_{\oplus}\sin{\theta_1}}{L_{\nu_{1_{w}}}}} \left(
1+\frac{2R_{\oplus}\sin{\theta_1}}{L_{\nu_{1_{w}}}} \right)\right]
\]

For the following ones the corresponding volumes are given by

\[
\frac{V_{2}(E_{\tau})}{A}=\frac{l_{\tau}}{\left(1-\frac{l_{\tau}}{%
L_{\nu_{1w}}}\right)} \left(1-e^{-\frac{L_0}{l_{\tau_2}}}\right)\int^{%
\theta_2}_{\theta_1}{d\theta} \sin{\theta}\cos{\theta}\cdot{e}^{-2R_{\oplus}%
\left[\left(\frac{\sin{\theta}-\sqrt{\cos^{2}{\theta_1}- \cos^2{\theta}}}{%
L_{\nu_{water}(E)}}\right)+\left(\frac{\sqrt{\cos^{2}{\theta_1}- \cos^2{%
\theta}}}{L_{\nu_{rock_1}(E)}}\right)\right]}
\]

\begin{eqnarray}
\frac{V_{3}(E_{\tau})}{A} & = & \frac{l_{\tau}}{\left(1-\frac{l_{\tau}}{%
L_{\nu_{1w}}}\right)} \left(1-e^{-\frac{L_0}{l_{\tau_2}}}\right)\int^{%
\theta_3}_{\theta_2}{d\theta} \sin{\theta}\cos{\theta}\cdot  \nonumber \\
& \cdot & {e}^{-2R_{\oplus}\left[\left(\frac{\sin{\theta}-\sqrt{\cos^{2}{%
\theta_1}- \cos^2{\theta}}}{L_{\nu_{water}(E)}}\right)+\left(\frac{\sqrt{%
\cos^{2}{\theta_1}-\cos^2{\theta}}- \sqrt{\cos^{2}{\theta_2}-\cos^2{\theta}}%
}{L_{\nu_{rock_1}(E)}}\right)+\left(\frac{\sqrt{\cos^{2}{\theta_2}- \cos^2{%
\theta}}}{L_{\nu_{rock_2}(E)}}\right)\right]} \nonumber
\end{eqnarray}

\begin{eqnarray}
\frac{V_{n}(E_{\tau})}{A} = \frac{l_{\tau}}{\left(1-\frac{l_{\tau}}{L_{\nu_1}%
}\right)} \left(1-e^{-\frac{L_0}{l_{\tau_2}}}\right)\cdot  \nonumber
\end{eqnarray}

\begin{eqnarray}
\cdot\int_{\theta_{n-1}}^{\theta_{n}}d\theta\sin(\theta)\cos(\theta)\exp%
\left\{ -2R_{0}\left[ \frac{\sin(\theta)}{L_{\nu_{_{1}}}}+\sum_{j=1,n-1}%
\sqrt{_{\cos^{2}(\theta_{j})-\cos^{2}(\theta)}}\left(\frac{1}{%
L_{\nu_{_{j+1}}}}-\frac{1}{L_{\nu_{_{j}}}}\right)\right] \right\}
\end{eqnarray}

where again, $\rho_{r_j}$ is the relative density of the matter
respect to the water, $L_{\nu_j} = (n_{\rho_{r_j}}
\sigma_{CC})^{-1}$, and
$\theta_j\equiv\arccos\frac{R_j}{R_{\oplus}}$.

Thus the total effective volume is

\[
\frac{V_{tot}(E_{\tau})_{Shells}}{A}=\sum_{n=1,N}\frac{V_n(E_{\tau})}{A}%
=\sum_{n=1,N}\frac{l_{\tau}{\left(1- e^{-\frac{L_0}{l_{\tau_2}}}\right)}}{%
\left(1-\frac{l_{\tau}}{L_{\nu_1}}\right)}\cdot
\]

\[
\cdot\int_{\theta_{n-1}}^{\theta_n}\sin({\theta} )\cos(\theta)\exp \left\{
-2R_{0}\left[ \frac{\sin(\theta)}{L_{\nu_{_{1}}}}+\sum_{j=1,n-1}\sqrt{%
_{\cos^{2}(\theta_{j})-\cos^{2}(\theta)}}\left(\frac {1}{L_{\nu_{_{j+1}}}}-%
\frac{1}{L_{\nu_{_{j}}}}\right) \right] \right\}
\]

This result coincides with the analytic expression of the
effective volume we have derived in the text using the
$D(\theta)$. In this way we have calibrated  the two methods and
we have verified that they converge to a unique result when we
calculate the effective volume, mass and event rate.

\begin{table}[htbp]
\begin{center}
\begin{tabular}{|c|c|c|}
\hline
&  &  \\
&  & $R_{j+1}(Km)>R>R_j(Km)$ \\
&  &  \\ \hline
&  &  \\
$L_{\nu_{1_{w}}} = L_{\nu_{water_1}}$ & $\rho_r=\,\, 1.02$ & $6371>R>6368$
\\
$L_{\nu_{1_{r}}}\,= L_{\nu_{rock_1}}\,\,$ & $\rho_r=\,\, 2.65$ & $6371>R>6368
$ \\
$L_{\nu_{2_{r}}}\,=L_{\nu_{rock_2}}\,\,$ & $\rho_r=\,\, 2.76$ & $6368>R>6346$
\\
$L_{\nu_{3_{r}}}\,=L_{\nu_{rock_3}} $ & $\rho_r=\,\, 3.63$\, & $6346>R>5700$
\\
$L_{\nu_{4_{r}}}\,=L_{\nu_{rock_4}}\,\,$ & $\rho_r=\,\, 5.05$\, & $%
5700>R>3480$ \\
$L_{\nu_{5_{r}}}\,=L_{\nu_{rock_5}}\,\,$ & $\rho_r=11.28$ & $3480>R>1221$ \\
$L_{\nu_{6_{r}}}\,=L_{\nu_{rock_6}}\,\,$ & $\rho_r=12.99$ & $1221>R>0$%
\,\,\,\,\,\,\,\,\, \\
&  &  \\ \hline
\end{tabular}
\end{center}
\end{table}
\

\subsection{Acknowledgment}

The author wish to thank Prof. Livio Scarsi for inspiring the
present search as well as the EUSO collaboration for the exciting
discussion; the author thanks  M.Teshima for technical
suggestions.

\bibliographystyle{plain}
\bibliography{xbib}

\end{document}